\documentclass[a4paper,12pt]{article}
\usepackage{jheppub}

\hypersetup{
    colorlinks,
    linkcolor={red!50!black},
    citecolor={blue!50!black},
    urlcolor={blue!80!black}
  }
      
\usepackage[T1]{fontenc}
\DeclareSymbolFont{usualmathcal}{OMS}{cmsy}{m}{n}
\DeclareSymbolFontAlphabet{\mathcal}{usualmathcal}

\usepackage{graphicx}
\usepackage{hyperref}

\usepackage{tensor}
\usepackage[most]{tcolorbox}
\usepackage{slashed}
\usepackage{bm}
\usepackage{braket}

\def\ra{\rightarrow}
\def\lra{\leftrightarrow}

\newcommand{\numberset}{\mathbb}

\newcommand{\R}{\numberset{R}}

\newcommand{\Z}{\numberset{Z}}

\newcommand{\p}{\partial}
\newcommand{\M}{\mathcal{M}}
\renewcommand{\d}{\mathrm{d}}

\def\be{\begin{equation}} 
\def\ee{\end{equation}}

\def\dag{\dagger} 
\def\a{\alpha} 
\def\b{\beta} 
\def\g{\gamma} 
 
\def\D{\Delta} 
\def\de{\delta} 
 
\def\eps{\varepsilon} 
\def\z{\zeta}

\def\l{\lambda} 
\def\L{\Lambda}

\def\r{\rho} 
\def\s{\sigma} 
\def\S{\Sigma} 
\def\t{\tau} 
 
\def\vf{\varphi}

\def\W{\Omega} 
\def\Th{\Theta} 
\def\th{\theta}
\def\wt{\widetilde} 
\def\wh{\widehat} 

\numberwithin{equation}{section}

\title{Bosonization of 2+1 dimensional fermions on the surface of topological
  insulators}
\author[a,1]{Andrea Cappelli,\note{Corresponding author.}}
\author[b,c]{Lorenzo Maffi,}
\author[d]{Riccardo Villa}

\affiliation[a]{INFN, Sezione di Firenze, Via G. Sansone, 1, 50019
  Sesto Fiorentino, Firenze, Italy}

\affiliation[b]{Dipartimento di Fisica e Astronomia G. Galilei, Universit\`a
  di Padova, I-35131 Padova, Italy}

\affiliation[c]{INFN, Sezione di Padova, I-35131 Padova, Italy}

\affiliation[d]{Dipartimento di Fisica,
  Universit\`a di Firenze, and INFN, Sezione di Firenze, Via
  G. Sansone, 1, 50019 Sesto Fiorentino, Firenze, Italy}

\emailAdd{andrea.cappelli@fi.infn.it}
\emailAdd{lorenzo.maffi@unipd.it}
\emailAdd{riccardo.villa@unifi.it}

\abstract{ Three-dimensional topological insulators can be described
by an effective field theory involving two `hydrodynamic' Abelian gauge
fields.  The action contains a bulk topological BF term and a
surface term, called loop model. This describes the massless 2+1 dimensional
excitations and provides them with a semiclassical,
yet non-trivial conformal invariant dynamics.
Given that topological insulators are originally fermionic, this physical
setting is ideal for realizing the bosonization of 
massless fermions in terms of gauge fields.
Building on earlier analyses of the loop model,
we find that fermions belong to the solitonic spectrum and can be described 
by Wilson lines, through the generalization of 1+1 dimensional vertex operators.
Their correlation functions agree with conformal invariance.
The bosonic loop model is then mapped into a fermionic theory by
using the general construction
of fermionic topological phases described in the literature.
It requires the identification of the characteristic
 one-form $\Z_2$ symmetry of the bosonic theory and its gauging, which
originates the fermion number $(-1)^F$, the spin sectors and the time
reversal symmetry obeying ${\cal T}^2=(-1)^F$.
These results are detailed for the effective action
and the partition function on the geometry $S^2\times S^1$.}

\begin{document}
\maketitle
\flushbottom

%-1------------------------------------------
\section{Introduction}

In this work we analyze the effective field theory of 3+1 dimensional
topological insulators \cite{wen2017colloquium,ryu2016classification},
which is based on the BF topological theory, with bulk and surface
terms \cite{BFTI}. In addition, the boundary dynamics is given by the
non-local Abelian gauge theory called loop model, which involves two
gauge fields, respectively `electric' and `magnetic' \cite{loopmodel}.
The loop model possesses several interesting features: it is conformal
invariant with a critical line, analogous to the compact boson in 1+1
dimensions; it is invariant under 2+1 dimensional particle-vortex
duality; and it matches both the selfdual mixed-dimension $QED_{4,3}$ for large
number of flavors $N_f\to\infty$ and the three-dimensional $QED$ in the
semiclassical limit.

In an earlier work by some of the authors, the loop model was quantized
and the partition function was obtained on the generalized cylinder
$S^2\times S^1$, giving access to the spectrum of conformal dimensions
of solitonic states \cite{loopmodel}.
In the present study, the corresponding field operators are
described in terms of open Wilson loops of the two
Abelian fields, which are analogous to order and disorder fields
\cite{frohlich95,marino}.
Their two-point functions are found to be conformal invariant
with scaling dimensions matching the solitonic spectrum in the partition
function.

The effective field theory description of topological phases
is a natural playground for understanding bosonization in higher dimensions,
mirroring what happens in the quantum Hall effect, where the
2+1d Chern--Simons theory in the bulk matches the 1+1d boundary
conformal theory bosonizing Weyl fermions  \cite{cappellianomaliescondmat}.
In one dimension higher, the bulk-boundary correspondence again
provides a guide for defining the field operators, their quantum
numbers and statistics, starting from the bulk topological theory and
then extending them to the boundary, where they acquire the
semiclassical solvable dynamics given by the loop model. Research on
bosonization in higher dimensions has a long history
\cite{marino,frohlich95,fradkin2017disorder} and recently received a
lot of attention
\cite{tongwebofduality,wittenwebofduality,kapustinExactbosonization,fradkin2018loopmodel}.

The fermion excitation is described in the loop model by a pair of order and disorder fields, whose charges requires a modification of the soliton spectrum, allowing half-integer flux quantization of gauge fields.

It turns out that the modified quantization  is one element
of the general construction
relating bosonic topological theories to fermionic `spin topological theories'
\cite{dijkgraafwitten,witten2016fermion}.
The first step is the identification of the one-form $\Z_2$ symmetry 
in the 2+1d bosonic theory generated by fermionic loops,  
with characteristic universal anomaly.
After its compensation by a bulk term, as usual in topological phases,
this symmetry is gauged
by introducing a dynamical $\Z_2$-valued two-form flat field $B$
\cite{gaiottokapustinspinTQFT1,gaiottokapustinspinTQFT2,
kapustinExactbosonization}.
This new field allows for half--integer flux quantization
and, furthermore, generates the `dual' $\Z_2$ fermionic parity symmetry. The spin sectors and the time-reversal
symmetry of the theory also follow.

In the second part of this work, we work out the detailed map between bosonic to fermionic theories, using the partition function
and the effective action, both including the loop model dynamics.
This is simple enough to allow several results at the topological
level to extend to the full theory in very explicit terms.

The paper is organized as follows. In section \ref{sec-2}, we recall
the topological BF description of the topological insulator and the
properties of the loop model \cite{loopmodel}.  In particular, the
soliton charges and statistics, the selfduality and the expression of
the partition function giving the scaling dimensions of soliton
fields.

In section \ref{sec-3}, we write
the operators that create soliton excitations, both in the
bulk and in the boundary. The corresponding open Wilson lines and surfaces
possess charges and statistics
determined by the topological theory. Next, we compute the
correlation functions of soliton operators. Even if the loop model
is a quadratic theory, the computation is not trivial: we rely on
two methods, the direct perturbative calculation and the nonperturbative prescription for disorder fields of Refs. \cite{frohlich87,frohlich95}
(details of the calculation are reported in appendix \ref{app-A}).

In section \ref{sec-4}, we focus on the construction of
the fermionic theory. It requires a modification of
the loop model by the introduction of the
dynamical field $B$. This couples to the current of the one--form
$\Z_2$ symmetry necessary for bosonization 
\cite{gaiottokapustinspinTQFT1,gaiottokapustinspinTQFT2,
  kapustinExactbosonization}.
Several aspects of the fermionic partition function are discussed, its spectrum,
spin sectors, and particle-vortex duality.
The additional, one--form
gauge invariance imposes constraints on the physical states.
The realization of fermion number and time reversal symmetries
is also described.

Finally, we show that two bosonic theories, with BF coupling $k=1$ and
also $k=2$, can be equivalently mapped into the same fermionic theory.
In the conclusion, we further comment on the difference between bosonic and
fermionic topological phases of matter and discuss developments
of our approach to bosonization in higher dimensions.

%-2------------------------------------------------
\section{Bosonic effective field theory of topological
  insulators}
\label{sec-2}

\subsection{Bulk topological theory}

In this section we review the description of (fractional) topological
insulators developed in previous works \cite{cappellirandellinisisti,loopmodel}. It is based on the bulk topological BF action
\cite{BFTI} together with the conformal invariant boundary dynamics
given by the loop model \cite{fradkin1996loopmodel,fradkin2018loopmodel}.

The degrees of freedom in the bulk are supposed to be point
particles and vortex lines, associated to 
currents $J^\mu$ and $J^{\mu\nu}$, respectively.
Being conserved, they can be parameterized as
\begin{equation}
  \label{curr-def}
  J^{\mu} = \frac{1}{2\pi} \epsilon^{\mu \nu\rho \sigma}
  \partial_\nu b_{\rho\sigma},
  \qquad\qquad
  J^{\mu \nu} = \frac{1}{2\pi} \epsilon^{\mu \nu\rho \sigma}
  \partial_\rho a_{\sigma},
\end{equation}
in terms of two Abelian $U(1)$ 'hydrodynamic' gauge fields, the one--form
$a = a_\mu \d x^\mu$ and the two--form
$b = \frac{1}{2} b_{\mu \nu} \d x^\mu \wedge \d x^\nu$.

The bulk action is 
\begin{equation}
\label{bulk-action}
S_{bulk}[b,a, A] = i\int_{\M_4} \frac{k}{2 \pi} b \d a +
\frac{1}{2\pi} b\d A - \frac{\theta}{8\pi^2} \d a \d A \,,
\end{equation}
where $A$ is the background electromagnetic field. 

The action (\ref{bulk-action}) correctly reproduces the basic features
of topological insulators.
Indeed, it is time--reversal (TR) invariant by requiring the
parameter of the third term to be $\theta= \pi\sim -\pi$, being
$\theta=0$ for trivial insulators.  Upon integration of the
hydrodynamic matter fields $a,b$, one obtains the induced action
\begin{equation}
\label{response-action}
    S_{ind}[A] = i\frac{\theta}{8\pi^2 k} \int_{\M_4}  \d A\, \d A \,,
\end{equation}
which expresses the response of the system to background changes.  For
$k =1$ this is the standard electromagnetic response of topological
insulators which can also be derived from a bulk Dirac fermion in the
limit of large mass \cite{cappellianomaliescondmat}.  For integer
$k\neq 1$, the theory \eqref{bulk-action} describes interacting
electrons forming the so-called fractional topological insulators, in
analogy with the physics of the quantum Hall effect\footnote{ As in
  that case, the response action \eqref{response-action} involves the
  fractional coupling constant $1/k$ and thus is not well-defined
  globally \cite{witten2016threelectures}. }
\cite{cappellianomaliescondmat}.
The basic excitations are point-like and loop-like,
with fractional charge and fractional monodromy phases.

In order to see these features, we couple the $a, b$ fields to point particle moving along $\xi(t)$
and vortex line sources in $\xi(t,\sigma)$, by adding the following terms to \eqref{bulk-action}:
\begin{equation}
\label{bulk-sources}
\begin{split}
    &\D S_{bulk}= -i\int_{\M_4} a_\mu \mathcal{J}^\mu +
                \frac{1}{2}b_{\mu \nu} \mathcal{J}^{\mu \nu},
                \\
& \mathcal{J}^\mu = N_0\, \dot\xi^\mu \de^{(3)}(x-\xi(t)),\qquad
\mathcal{J}^{\mu\nu} = M_0\, \dot\xi^\mu\times \xi^{'\nu} \de^{(2)}(x-\xi(t,\s)), 
\end{split}
\end{equation}
with respective charges $N_0,M_0 \in \Z$.  The equations of motion for
the $a,b$ fields obtained from the action \eqref{bulk-action},
\eqref{bulk-sources} and the definitions \eqref{curr-def} imply
\begin{equation}
\label{bulk-em}
\begin{split}
&J^\mu \equiv \frac{1}{2\pi} \eps^{\mu\nu\a\b}\p_\nu b_{\a\b}=
\frac{1}{k}{\cal J}^\mu,
  \\
& J^{\mu\nu}\equiv \frac{1}{2\pi} \eps^{\mu\nu\a\b}\p_\a a_\b=
\frac{1}{k}{\cal J}^{\mu\nu} - \frac{1}{2\pi} \eps^{\mu\nu\a\b}\p_\a A_\b\,.
\end{split}
\end{equation}
The currents can be integrated across surfaces $\S$ and volumes
${\cal V}$ encircling the lines and points, respectively, leading to the following bulk charges 
\begin{equation}
\label{bulk-charges}
\begin{split}
{\cal Q} &=\int_{\cal V} J^0=
\frac{1}{2 \pi} \int_{\cal V} \d b = \frac{N_0}{k},
\\
{\cal Q}_T &=\int_\S  J^{0i} n^i =
     \frac{1}{2 \pi} \int_{\S} \d a =\left( \frac{M_0}{k} -
     \frac{1}{2 \pi} \int_{\S} \d A \right),
\end{split}
\end{equation}
where $n^i$ is the unit vector orthogonal to the surface $\S$.
These equations shows that both excitations have charges in units of
$1/k$, as expected. The second equation expresses the charge density
along the vortex line, whose aspects will be better described
later.

The main feature of the BF action is to attach fractional fluxes to
charges, leading to Aharonov-Bohm phases for point particles
encircling vortex lines in three-dimensional space \cite{loopmodel},
\be
\frac{\Th}{2\pi}=\frac{N_0 M_0}{k}\,.
\label{bulk-phase}
\ee

%-2.2------------------------------------------
\subsection{Boundary theory: the loop model}

Topological insulators possess (interacting) massless excitations at
their surface. The bulk theory \eqref{bulk-action}
determines the type of boundary fields and the topological part of the
boundary action (no Hamiltonian), as follows
\begin{equation}
\label{top-surface-action}
S_{surf}[\zeta, a, A] = i \int_{\M_3} \frac{k}{2\pi} \zeta \d a +
\frac{1}{2 \pi} \zeta \d A -\frac{1}{8\pi} a\d A,
\end{equation}
where $a$ is the bulk field brought to the boundary
$\M_3=\p \M_4$ and $\z$ is the reduction to the boundary of 
the two-form, $b|_{{\cal M}_3}= \d\z$.
The whole action involving bulk and
boundary terms \eqref{bulk-action}, \eqref{top-surface-action},
is invariant under local gauge transformations,
$b \ra b + \d \beta$, $\zeta \ra \zeta + \beta$ and $a\ra a+d\vf$.

The last term in the action, once the $a,\z$ fields are integrated out,
determines the boundary response, given by the Chern-Simons action
$\int A\d A$ with coupling constant $1/2k$. For $k=1$, this the
well-known parity (and time-reversal) global anomaly of 2+1
dimensional fermions, which is canceled by the theta term in the bulk
action \eqref{bulk-action}, according to the bulk-boundary
correspondence.\footnote{ A deeper description of global anomaly
  cancellation in three-dimensional topological insulators can be
  found in \cite{witten2016fermion} and is reviewed in section 10 of
  \cite{cappellianomaliescondmat}.}  In effective field theory
approaches, the anomaly is described at classical level: in the
present case, it amounts to a term in the action added by hand, with
natural generalization to $k>1$ \cite{cappellianomaliescondmat}.  As a consequence of anomaly
cancellation, the boundary anomaly in \eqref{top-surface-action}
and the bulk theta term in \eqref{bulk-action} are sometimes
omitted in the following discussion.

The boundary action \eqref{top-surface-action} should be
completed by adding a dynamical term for massless excitations. The
proposal by Ref. \cite{fradkinloop,loopmodel} is to introduce the
following non--local quadratic dynamics for $a_\mu$
\begin{equation}
\label{loop-model-dyn}
S_{dyn}[a] = \frac{g_0}{4\pi} \int \d^3 x \d^3 y\, a_\mu(x)
\frac{-\delta_{\mu \nu} \partial^2 + \partial_\mu \partial_\nu}
{\partial } a_\nu (y),
\end{equation}
where $g_0$ is a coupling constant, $\p^2 = \p_\mu \p_\mu$
and $\partial^{-1}$ is the Green function of
$\partial \equiv \sqrt{-\partial^2}$. Explicitly in three Euclidean
dimensions:
\begin{equation}
    \frac{1}{\p (x,y)}=\frac{1}{2 \pi^2} \frac{1}{(x-y)^2}.
\end{equation}
 All in all, the whole action for the boundary degrees of freedom is\footnote{
Hereafter omitting the subscript to $S_{surf}$.}
   \begin{equation}
\label{loop-model}
\begin{split}
  S[\zeta, a, A] &= \frac{i}{2\pi}\int_{\M_3}   k \zeta \d a +
    \zeta \d A  +   \frac{g_0}{4\pi} \int_{\M_3} a_\mu
  \frac{-\delta_{\mu \nu} \partial^2 + \partial_\mu \partial_\nu}
  {\partial} a_\nu =
  \\
  &= \frac{i}{2\pi}\int_{\M_3}  k \zeta \d a +  \zeta \d A  +
  \frac{g_0}{8\pi} \int_{\M_3} f_{\mu\nu} \frac{1}{\partial}  f_{\mu\nu},
\end{split}
\end{equation}
where $f_{\mu\nu}$ the field strength of the $a_\mu$ field.
This theory is called the `loop model', because it can also be written
in terms of closed loops, \textit{i.e.}$\,$ currents $j=*\d a$, interacting
with $1/(x-y)^2$ potential \cite{fradkin1996loopmodel}.
This kind of dynamics was
also considered in \cite{zirnstein2013cancellation,fradkin2018loopmodel}.
The loop model \eqref{loop-model} is the main focus of the present work.

The integration of the field $\zeta$ implies $a = -A/k$ and
leads to the induced action
\begin{equation}
\label{loop-ind-action}
S_{ind}[A] = \frac{g_0}{4\pi k^2} \int_{\M_3}  A_\mu
\frac{-\delta_{\mu \nu} \partial^2 + \partial_\mu \partial_\nu}{\partial} A_\nu.
\end{equation}
This is the same response of 2+1d fermions to quadratic order
$O(A^2)$, for $k=1$ and $g_0 = \pi /8$, having canceled the parity anomaly as said.
Thus, the loop model reproduces a feature of the
underlying fermionic theory at the microscopic level. The
approximation holds for $\hbar\to 0$ or weak fields $A_\mu$:
equivalently, Eq.\eqref{loop-ind-action}
is the leading term for large number of flavors $N_f\to\infty$.
Such scale-invariant response provides a good reason for considering
this nonlocal dynamics at the boundary of topological insulators
\cite{loopmodel}.

The kernel of the loop model satisfies the following
nice inversion property
\begin{equation}
\label{kernel-loop-model}
\begin{split}
 \int j_\mu D_{\mu\nu}^{-1} j_\nu & = \int \z_\mu D_{\mu\nu} \z_\nu,
 \\
 D_{\mu\nu}(x,y) & =\frac{-\delta_{\mu \nu} \partial^2 +
   \partial_\mu \partial_\nu}{\partial}(x,y),
 \qquad\qquad j_\mu = \epsilon_{\mu \nu \rho}\p_\nu \z_\rho.
\end{split}
\end{equation}
Using this identity we can integrate out the field $a$ in the action
\eqref{loop-model} and obtain the following expression involving the
$\zeta$ field only:
\begin{equation}
\label{loop-model-zeta}
S[\zeta, A] = \frac{k^2}{4\pi g_0} \int_{\M_3}
\zeta_\mu \frac{-\delta_{\mu \nu} \partial^2 +
  \partial_\mu \partial_\nu}{\partial} \zeta_\nu +
\frac{i}{2\pi} \int_{\M_3} \zeta \d A.
\end{equation}

%-2.3------------------------------------------
\subsection{Properties of the loop model}

In the following, we list some properties of this 2+1 dimensional
theory discussed in \cite{loopmodel}, focalizing on the geometry of
the generalized cylinder $S^2\times \R$, with compact space and linear
time, where the topological insulator bulk is the interior of the ball $D^3$
of radius $R$, $D^3=S^2\times I$, $I\in [0,R]$.

\subsubsection{Soliton charges}

\begin{figure}
 \begin{center}
  \includegraphics[scale=.18]{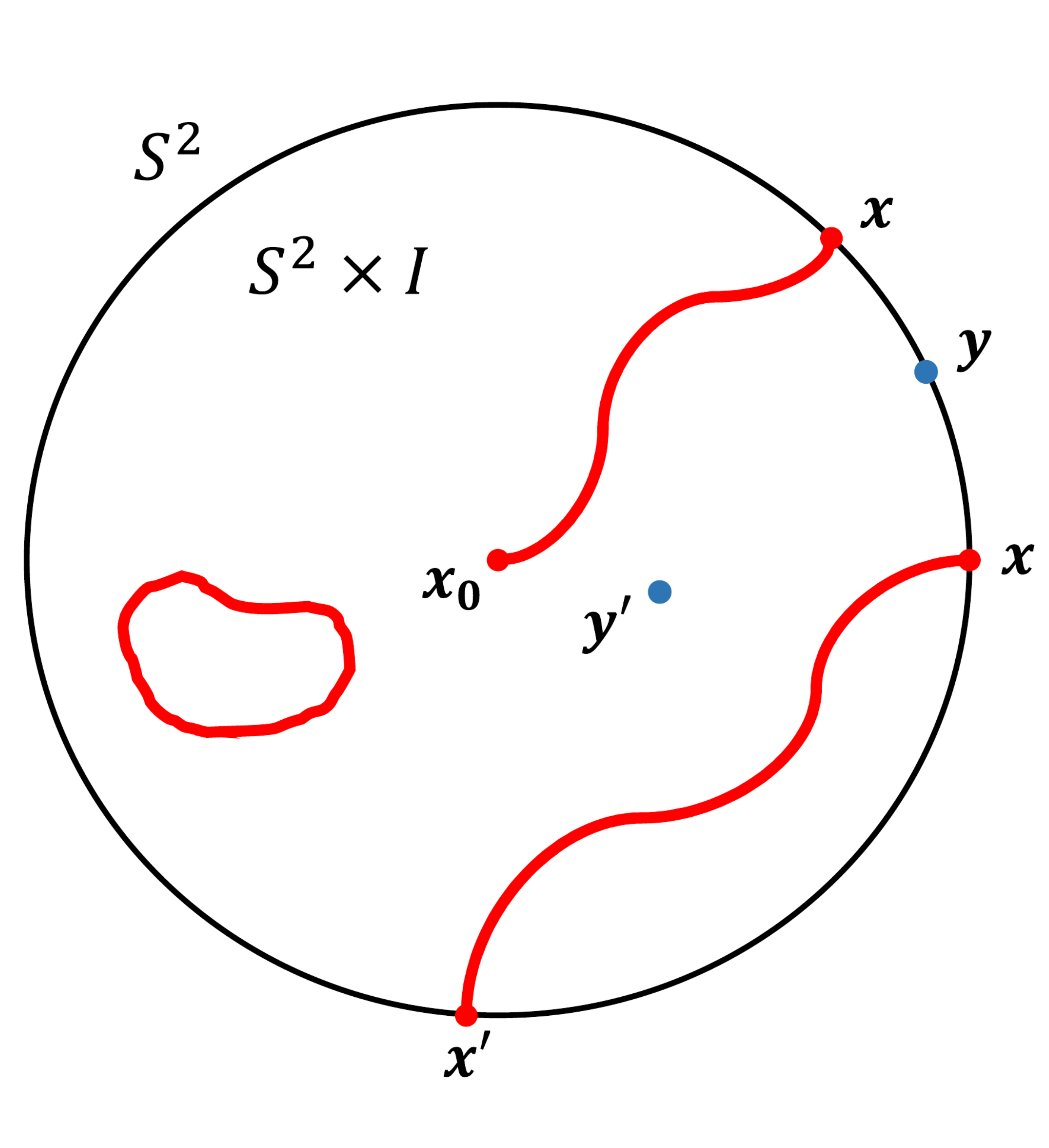}
 \end{center}
 \caption{The spatial geometry with bulk $D^3=S^2\times I$ and
   boundary $S^2$. The vortex lines are drawn in red, they can finish
   on the boundary at points $x,x'$, where they become point sources, or at the singular center $x_0$. A
   closed vortex line is also shown.  Point sources at $y,y'$ of the other field
   are in blue, in the bulk and at the boundary.}
 \label{fig1}
\end{figure}

A standard fact of topological matter is that bulk excitations correspond
to solitonic sectors of the boundary theory, with associated conserved
charges. Let us see how the flat bulk connections $a,b$ with
characteristic quantized fluxes
\eqref{bulk-charges} extend to the boundary, where $b=\d \z$.
A bulk point source generates the flux $\int \d\z$ across the
boundary, which remains invariant when the source is brought
to the boundary.
The bulk vortex line can be open or closed; in the open case, it hits the
boundary once or twice, with endpoints becoming another kind of
point-like excitation with flux $\pm\int \d a$. The line  can end
inside the ball at the singular center, which can be thought of as
the zero-radius limit of an inner boundary sphere, not considered
here for simplicity (see Fig.\ref{fig1}).
A closed vortex line can be fully inside the bulk or lay on the boundary.

The fluxes associated to point-like excitations in the boundary
are therefore
\begin{equation}
\label{boundary-charges}
Q= \frac{1}{2\pi}\int_{S^2} \d \zeta  = \frac{N_0}{k},
\qquad\qquad Q_T = \frac{1}{2 \pi} \int_{S^2} \d a  = \frac{ M_0}{k}.
\end{equation}
The corresponding solitons of the theory (\ref{loop-model}) are
solutions of the equations of motion with  boundary conditions suitable
for these charges.

Note that the $\z$ flux characterizes a soliton with electric charge
$Q$, owing to the original coupling to the electromagnetic background
in \eqref{loop-model}, thus the field $\z$ is `magnetic'. The field
$a$ is an ordinary `electric' gauge field like $A$, such that the
corresponding solitons have magnetic charge $Q_T$ (historically called
`topological' charge \cite{marino}).

%-2.3.2------------------------------------------
\subsubsection{Self--duality }
\label{sec-duality}

Particle-vortex duality in 2+1 dimensions is, roughly speaking,
equivalent to switching the
coupling to the $A$ background from electric to magnetic \cite{tongwebofduality,wittenwebofduality}. This is
achieved by replacing $A_\mu$ with the auxiliary
dynamical field $k c_\mu$, which itself couples to $A$ by a
mixed Chern-Simons term.
The original and dual theories, respectively $Z[A]$ and
$\wt Z[A]$, are thus related by
\be
\wt Z[A]=\int {\cal D}c_\mu\, Z[k c_\mu] \exp\left(i\frac{1}{2\pi}\int c\,
  \d A\right),
\ee
which is a kind of quantum analog of the Legendre transform \cite{loopmodel}.
Upon insertion of the induced action \eqref{loop-ind-action} for
$Z[k c_\mu]$ and integration, one finds that the loop model is
selfdual, with the following map between couplings, fields and charges
\begin{equation}
\label{loop-model-duality}
g_0\ \lra\ \wt g_o=\frac{k^2}{g_0}, \qquad
\z_\mu \ \lra\ \wt\z_\mu =a_\mu, \qquad  N_0 \ \lra\ \wt N_0= M_0 .
\end{equation}
These relations can be proven by introducing the coupling to
the $k c_\mu$ field in the action \eqref{loop-model}:
integrating over $\z_\mu$ and solving for $c_\mu$ leads to the expression
\begin{equation}
\label{dual-loop-model} 
\wt S[a, A] =  \frac{g_0}{4\pi}
\int a_\mu \frac{-\delta_{\mu \nu} \p^2 + \p_\mu \p_\nu}{\p} a_\nu +
\frac{i}{2\pi} \int a \d A\,,
\end{equation}
which should be compared to \eqref{loop-model-zeta}.
The correspondence between the charges $(N_0,M_0)$ confirms that
the monodromy phase \eqref{bulk-phase} is selfdual, as expected.

Note that the factor of $k$ in the duality transformation
\eqref{loop-model-duality}, extending the map considered in the
literature \cite{tongwebofduality}, is rather natural for charges and monodromies,
which should be independent of the field-theory description.

%-2.3.3------------------------------------------
\subsubsection{Phase diagram}

The loop model possesses a critical line parameterized by the coupling
$g_0$, where $g_0 \ge k$ can be chosen owing to its self-duality. This
result was obtained by numerical simulations \cite{motrunich2012} and
is confirmed by Peierls arguments \cite{loopmodel}. It is
found that the monopole condensation of $QED_3$ \cite{POLYAKOV1977} is
absent in the loop model, due to the non-local interaction: one is
left with a Coulomb-like phase, analogous to that occurring in 1+1d XY
model (the $c=1$ compact boson conformal field theory).

Another interesting fact is that the loop model can be shown to
correspond to the large $N_f$ limit of mixed-dimension $QED_{4,3}$,
involving 2+1d fermions and 3+1d photons \cite{loopmodel}.  This
theory is exactly particle-vortex self-dual and possesses a critical
line \cite{sonmixedqed,sonmixedqedscalar,dudal2019exact}. The direct
relation with this fermionic theory strongly suggests that 2+1
dimensional bosonization can be understood in the present framework.

%-2.3.4------------------------------------------
\subsubsection{Solitonic spectrum and conformal invariance}
\label{loop-model-CFT}

The partition function of the loop model has been obtained on the the
generalized cylinder $S^2 \times \R$, with radius $R$ and time $u$ 
\cite{loopmodel}.
The conformal map $r = R \exp (u/R)$ with flat Euclidean space,
where $r^2=(x_\mu)^2$, implies that time translations on the cylinder
correspond to dilatations in $\R^3$. Thus, canonical quantization on the
cylinder determines energy levels which corresponds to
scaling dimensions $\D$ of fields in the conformal theory,
schematically,
\begin{equation}
\label{CFT-Z}
    Z = \sum_\Delta e^{- \beta\Delta/R}.
\end{equation}

The quantization requires a Hamiltonian formulation which is
actually nontrivial due to the 
non--local nature of the action \eqref{loop-model}. It was obtained
by rewriting the loop model as a local theory in four
dimensions (with the extra dimension just a fictitious one),
leading to the result\footnote{
  Clearly, the partition function is computed on
  the periodic time interval $L=\b=1/k_B T$, thus the geometry is
  actually $S^2 \times S^1$.} 
\begin{equation}
\begin{split}
  \label{loop-model-Z}
  Z[S^2 \times S^1] &= Z_{osc}\, Z_{sol}= Z_{osc}\sum_{N_0,M_0 \in \Z}
  \exp \left( -\frac{\beta}{R} \Delta_{N_0,M_0} \right),
\\
   \Delta_{N_0,M_0} &= \lambda \left( \frac{N_0^2}{g_0} +
    g_0 \frac{M_0^2}{k^2} \right).  
\end{split}
\end{equation}
In this equation, $Z_{sol}$ is obtained by evaluating the classical action on
soliton configurations, leading to the spectrum of scaling dimensions
$\D_{N_0,M_0}$, and $Z_{osc}$ describes fluctuations around
classical solutions.
The soliton dimensions confirm the selfduality of the theory
\eqref{loop-model-duality}. The dimensionless parameter $\l=1/(MR)$ is
left undetermined, being related to the value of the infrared cutoff
$M$ needed in the calculation. It will be fixed later by computing
correlators of soliton fields (see section \ref{sec-solitons}).
The term $Z_{osc}$ has the form\footnote{This is actually the square of the generating
function of plane partitions or MacMahon function.} \cite{loopmodel}
\be
\label{Z-osc}
Z_{osc}= \frac{1}{\prod_{k=1}^\infty \left( 1- q^k\right)^{2k}},
\qquad\qquad q= e^{-\b/R}.
\ee
This expression is instructive
because it checks conformal invariance of the theory at the quantum
level: the Casimir energy vanishes, corresponding to no conformal
anomaly in 2+1d, and the dimensions $\D$ of descendant fields are
integer spaced.  These results confirm that the non-local nature of
the loop model causes no problem for conformal invariance.

%-2.3.5------------------------------------------
\subsubsection{Topological order}

The 3+1d BF theory \eqref{bulk-action} on the geometry
$D^3 \times S^1$ has $k$ degenerate ground states (topological order),
as easily obtained by standard methods (see \textit{e.g.} \cite{cappellirandellinisisti}). Therefore,
the partition function of boundary excitations \eqref{loop-model-Z}
should involve $k$ corresponding topological sectors.
They can be obtained by writing $M_0 = \ell + k m_0$, with $\ell\in\Z_k$
and $m_0 \in \Z$, as follows
\begin{equation}
  Z = \sum_{\ell=0}^{k-1} Z_\ell, \qquad\
  Z_\ell =Z_{osc} \sum_{N_0,m_0 \in \Z} \exp \left[ -\frac{\b\l}{R}
    \left( \frac{N_0^2}{g_0} + g_0 \left(  \frac{\ell}{k}+m_0 \right)^2
    \right) \right].
\end{equation}
A single anyonic sector $Z_\ell$, characterized by fractional charge
$M_0/k =\ell/k+m_0$, can be selected by placing an opposite
fractional charge in the interior of the sphere, and requiring integer
total charge in the system.  This is the standard choice for the
quantum Hall effect in the geometry of the spatial disk $D^2$ with a
single boundary circle $S^1$ \cite{cappelli1997}.

We remark that the loop model action could be modified by adding
further interactions consistent with conformal invariance; in these
cases, the coupling constant $g_0$ and dimensions $\D$ are
renormalized. For example, one could introduce three-body interactions
suggested by the $QED_{4,3}$ theory at order $O(1/N_f)$. Nonetheless,
the decomposition in $k$ sectors should stay, being a topological
property.  Many results on bosonization found later only depend on the
topological part of the bulk and boundary actions, and thus are valid
for more general dynamics. In this sense, the simple quadratic
interaction of the loop model can be considered as a reference case.

%-3------------------------------------------
\section{Soliton fields: topology and dynamics}
\label{sec-3}

In this section we construct the soliton fields, \textit{i.e.} the
operators charged under (\ref{bulk-charges}).  These are expressed by
Wilson lines (open loops) of the connections $a$, $\zeta$ occurring in the
boundary theory \eqref{loop-model}. We first write their expressions
and determine their charges and monodromy
phases by using the topological part of the action (\textit{i.e.} setting $g_0=0$);
we shall also see that these operators have counterparts in the bulk,
and discuss the bulk-boundary correspondence for them.
In the last subsection, we shall add the dynamical term in the
action ($g_0\neq 0$) and determine the correlation functions
finding agreement with conformal invariance.

The solitonic fields are first obtained in the case of
the 2+1 dimensional quantum Hall effect, thus `re-discovering' the
well-known bosonization of 1+1d relativistic fermion fields by vertex
operators. This sets the stage for the discussion of higher
dimensions.  We shall employ earlier approaches on solitons and
order-disorder fields \cite{marino,frohlich87}.

%-3.1---------------------------------------------
\subsection{Wilson lines and vertex operators in the QHE}
\label{sec-sigma-mu}

The effective theory of the quantum Hall effect at filling fractions
$\nu=1/k$ is expressed in terms of the Chern-Simons action, as follows
\cite{cappellianomaliescondmat}
\be
\label{FQHE}
S_{CS}[a,A]=i\int_{{\cal M}_3} \frac{k}{4\pi} a\d a +\frac{1}{2\pi} a
\d A - a_\mu {\cal J}^\mu\,.
\ee
This is analogous to the BF theory (see \eqref{bulk-action}), but
differ for the presence of a single hydrodynamic field
$a_\mu$. As a consequence, time--reversal invariance is absent in
\eqref{FQHE} and can only be achieved by considering two fields.

The action \eqref{FQHE} describes a massive topological phase in the
2+1d bulk: upon integrating over the $a$ field, one obtains (locally)
the response of the Hall current $J^i=1/(2\pi k)\eps^{ij} {\cal E}_j$,
$i,j=1,2$, where ${\cal E}_i$ is the in-plane electric field.

The Hamiltonian quantization of the theory is now considered in the
$a_0=0$ gauge and for vanishing background, $A=0$. The action just involves
the symplectic structure\footnote{Being first-order in time
derivatives, it is already written in Hamiltonian form
\cite{wittenNonAbelianBosonization,floreaninijackiw}.}
($H=0$) leading to the canonical commutators
\be
\label{CS-comm}
\left[a_i(t,\mathbf{x}),\pi_j(t,\mathbf{y}) \right]=
i\de_{ij} \de^{(2)} (\mathbf{x}-\mathbf{y}),
\qquad
\pi_i =\frac{k}{2\pi}\eps_{ij} a_j, \qquad i,j=1,2.
\ee

Given any canonical structure, one can define `electric'
and `magnetic' charges and write the field operators which create them,
as discussed in Ref.\cite{marino}.
The charges are written, with convenient normalizations,
\be
\label{FQHE-charges}
\mathcal{Q}[\pi]=\frac{1}{k}\int_{\S} \p_i \pi_i =\frac{1}{2\pi}\int_{\S}
\eps^{ij} \p_i a_j,
\qquad\qquad
\mathcal{Q}_T[a]=\frac{1}{2\pi}\int_{\S} \eps^{ij} \p_i a_j \,.
\ee
In this and following equations we first give the general expressions
in terms of $(a_i,\pi_j)$ and then their implementation in the
Chern--Simons theory \eqref{FQHE}, which realizes the simpler case $\mathcal{Q}=\mathcal{Q}_T$.

The open Wilson lines are defined by
\begin{align}
  \label{FQHE-sigma-mu}
  &\s(t,\mathbf{x}, \mathbf{x}_0 ) =
    \exp \left( -i\int_{\mathbf{x}_0,\g}^\mathbf{x}
    \d \xi^i a_i (t,\mathbf{\xi}) \right) =
    e^{-i \int^\mathbf{x}_\mathbf{x_0} a}, \\
    &\mu(t,\mathbf{x}, \mathbf{x}_0 ) =
    \exp \left( i\frac{2\pi}{k}\int_{\mathbf{x}_0,\g'}^\mathbf{x}
      \d \xi^i \eps_{ij}\pi_j(t,\mathbf{\xi}) \right) =
      \s(t,\mathbf{x}, \mathbf{x}_0 ).
\end{align}
These expressions involve spatial contours $\g,\g'$ but are actually
independent of their shape, because $a$ is flat: thus, they are
functions of the end-points only. They are defined at equal time
in the Hamiltonian formalism, but can be easily covariantized.
We remark that the names $\s,\mu$ are 
reminiscent of order-disorder fields in the Ising model, but actually have no
direct relation with them. Note that the simpler realization $\s=\mu$
is obtained for Chern--Simons.

The commutator of charges and Wilson lines is found by using a
Schr\"{o}dinger representation of commutation relations
\eqref{CS-comm}, with the result
\begin{equation}
\label{charge-comm}
\left[\mathcal{Q}, \sigma (t,\mathbf{x}, \mathbf{x}_0 ) \right] =
\s(t,\mathbf{x}, \mathbf{x}_0 ) \frac{1}{k}  \int_{\S}\d^2 y
\left[\de^{(2)}(\mathbf{y}- \mathbf{x}) -
\de^{(2)}(\mathbf{y}- \mathbf{x}_0) \right] . 
\end{equation}
Clearly, the Wilson lines create opposite fractional charges at their endpoints,
which can be selected by restricting the area of integration $\S$
in the definition of $\mathcal{Q}$ in \eqref{FQHE-charges}. Analogous
results are found for $[\mathcal{Q}_T,\mu]$.

\begin{figure}
 \begin{center}
  \includegraphics[scale=.12]{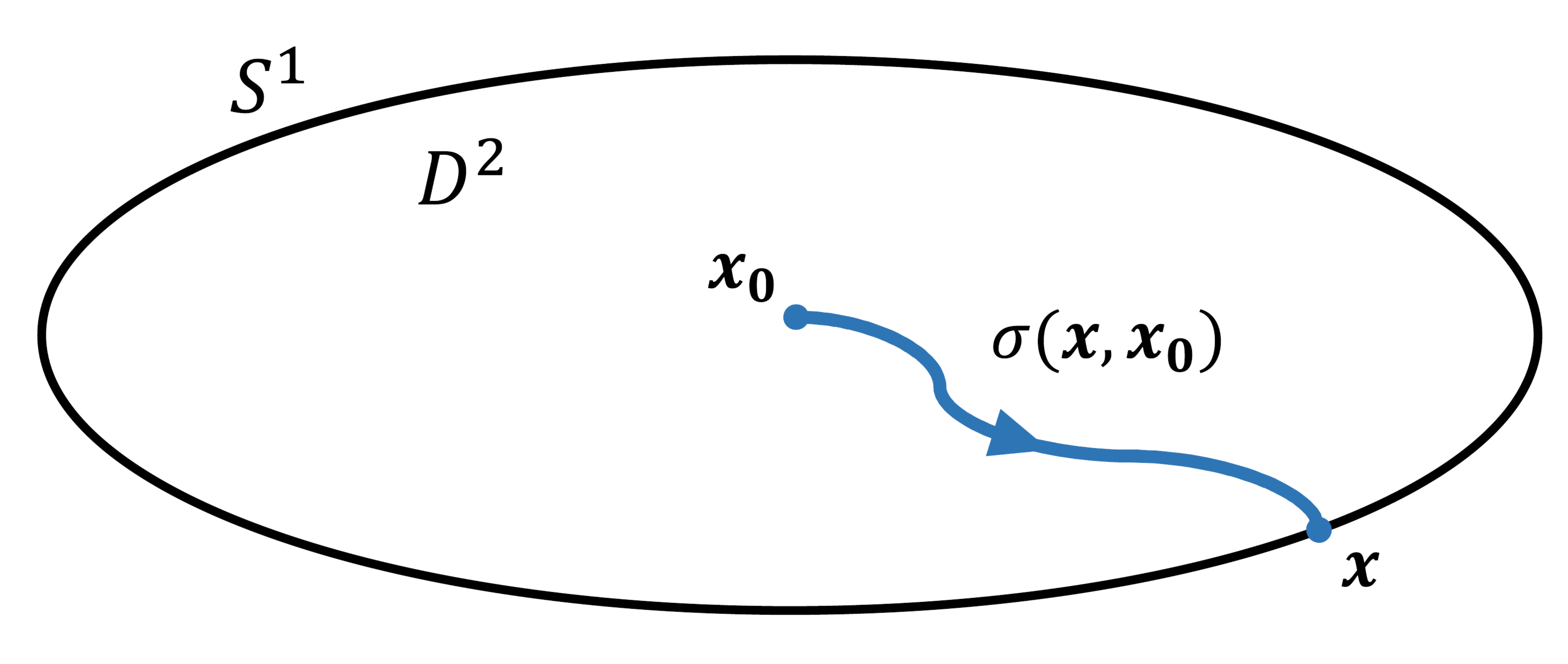}
 \end{center}
 \caption{Quantum Hall effect on the disk $D^2$. In blue, the
   Wilson line which ends on the boundary circle at $x$, where it becomes a
   pointlike field, the vertex operator.}
 \label{fig2}
\end{figure}

We recall that quantum Hall states possess massless excitations at the
edge of the system. Let us now show how the bulk line operators
\eqref{FQHE-sigma-mu} can be related to the fields living at the
boundary.  We consider the geometry of the disk $D_2$ with boundary
circle $S_1$ of radius $R$, for example (see Fig.\ref{fig2}).  The
boundary theory is obtained from bulk Chern-Simons
\cite{cappellianomaliescondmat} by
considering the gauge $a_0=0$ and reducing the flat gauge field to a
scalar function on the circle, $a_i|_{S^1}=\p_i \vf$, where $i=0,1$,
the circle coordinates being $(x^0,x^1)=(t,x=R\th)$.  The result is,
for $A=0$,
\be
\label{chiral-boson}
S[\vf]=-\frac{k}{4\pi}\int d^2x\; \p_x \vf \dot\vf + (\p_x \vf)^2 .
\ee
The first term in the action, the symplectic form, is obtained from
the edge reduction of Chern-Simons just discussed; the Hamiltonian part
$H\sim (\p_x\vf)^2$ is a further addition which gives dynamics to the edge
excitations. These are chiral waves, $\vf=\vf(t-x)$,
and soliton excitations due to the
compactness of the field $\vf$ \cite{cappellianomaliescondmat}.
The action \eqref{chiral-boson} defines the so-called chiral boson theory
\cite{floreaninijackiw},
a conformal theory with central charges $(c,\bar c)=(1,0)$
and a critical line parameterized by $k$ \cite{CDTZ}.
In particular, the Weyl fermion is found at $k=1$, thus establishing the
bosonization map.

The canonical structure is\footnote{The factor of two
  in the definition of the momentum is explained in \cite{floreaninijackiw}.}
\be
\label{chiral-boson-comm}
\left[ \vf(t,x), \pi(t,y)\right]=i \de(x-y),
\qquad\qquad \pi=-\frac{k}{2\pi} \p_x\vf,
\ee
and allows to define charges and operators proper of the boundary
theory
\begin{align}
  \label{chiral-boson-op}
  \begin{split}
  &  \wh Q[\pi]=-\frac{1}{k}\int_{S^1}dx\, \pi =
    \frac{1}{2\pi}\int_{S^1} dx\, \p_x \vf,
  \qquad\qquad \wh Q_T[\vf]= \frac{1}{2\pi}\int_{S^1}dx\, \p_x\vf\,,
  \\
  & \wh\s(t,x,x_0)=\exp\left(-i \vf(t,x)+i\vf (t,x_0) \right),
  \\
  & \wh\mu(t,x,x_0)=
    \exp\left(i \frac{2\pi}{k} \int_{x_0}^x d\xi \pi(t,\xi) \right)
    = \wh\s(t,x,x_0).
    \end{split}
\end{align}
These quantities are hatted not to confuse them with the bulk
expressions \eqref{FQHE-charges}, \eqref{FQHE-sigma-mu}. Note again
that charges and fields are equal in pairs, $\wh Q=\wh Q_T$,
$\wh \s =\wh \mu$ and obey commutation relations analogous to
\eqref{charge-comm} for points on $S^1$.

The following argument allows to neglect the dependence on the
reference point $x_0$ in $\wh \s(t,x,x_0)$. We suppose that
all operators have common point $x_0$:
if the total charge vanishes, then the dependence on $x_0$ cancels out.
If it does not, it leaves a factor $\exp(i Q_\infty \vf (t,x_0))$:
by conformal invariance of the theory, the reference point can be
sent to infinity in the future direction, along the cylinder
$S^1\times\R$ axis, where it modifies the
left ground state, leading to the so-called `charge at infinity'
of conformal field theory \cite{ginsparg1988applied}.
For example, we can rewrite the
two-point correlator in standard CFT notation as follows
\be
\lim_{|x^\mu_0|\to\infty} |x_0|^{2\D_{\wh\s}}
  \langle\W| \wh\s(x,x_0)\, \wh\s(y,x_0)|\W\rangle =
  \langle -2/k|e^{-i\vf(x)}\, e^{-i \vf(y)}|\W\rangle,
\ee
where $\hat{Q}_\infty=-2/k$ in this case.

Keeping in mind global charge conservation, we can omit the
$x_0$ dependence and identify the chiral vertex operator of the
conformal theory as  $\s(x)=\exp(-i\vf(x))$. 
The operator exchange on the line
\be
\label{boson-stat}
\s(t,x)\s(t,y)= e^{i\pi/k} \s(t,y)\s(t,x), \qquad\qquad x>y,
\ee
can also be obtained from the commutation relations \eqref{chiral-boson-comm},
integrated as
\be
[\vf (t,x),\vf (t,y)]=i (\pi/k) \eps(x-y), \ee where
$\eps(x)=\pm 1$ is the step function.

The result \eqref{boson-stat} identifies the Weyl fermion field
$\psi=\s$ for $k=1$.  One can show that the dynamics of the chiral
boson also implies the correct power-law correlator for
$\langle \psi \psi^\dag\rangle$ \cite{CDTZ}.

%-3.1.1------------------------------------
\subsubsection{Bulk-boundary correspondence for vertex operators}

Having discussed the definition of charged operators both in
the bulk and boundary of the QHE system, we can compare 
their expressions.
We clearly see that 1+1d vertex operators \eqref{chiral-boson-op} are basically the
2+1d Wilson lines \eqref{FQHE-sigma-mu} moved to points of the boundary circle,
where $a_i\to \p_i\vf$.

The fact that fermion fields can be written in terms of
Wilson lines of effective gauge fields $a,\pi$ is a key feature.
It shows that bosonization can be understood in the physical setting
of topological phases of matter, involving hydrodynamic topological actions
and the bulk-boundary correspondence.

%-3.2-----------------------------------------------
\subsection{Soliton fields on the boundary of 3+1d
  topological insulators}

The 2+1 dimensional fields creating point-like excitations with
charges $Q$, $Q_T$ in \eqref{boundary-charges} can be defined on the boundary of a topological insulator  by extending the
previous Hamiltonian approach to the BF action \eqref{loop-model}.
In the gauge $a_0=\z_0=0$, one finds the following commutation
relations involving the $(a,\zeta)$ fields
\begin{equation}
\label{BF-comm}
\left[ a_i(t, \mathbf{x}), \z_j (t, \mathbf{y}) \right] =
i\frac{2\pi}{k} \eps_{ij} \delta^{(2)} (\mathbf{x}-\mathbf{y}). 
\end{equation}
We now have two independent fields
\begin{equation}
\label{sigma-mu-op}
    \begin{split}
      & \sigma_{N_0}(\mathbf{x}) =\exp \left( -iN_0\int_{\mathbf{x}_0}^{\mathbf{x}}
        \d \xi^i\, a_i (t, \mathbf{\xi}) \right), \\
      & \mu_{M_0} (\mathbf{x})= \exp \left( -i M_0\int_{\mathbf{x}_0}^{\mathbf{x}}
        \d \xi^i \,\zeta_i (t, \mathbf{\xi}) \right) ,
\end{split}
\end{equation}
which obey the following relations with the charges \eqref{boundary-charges},
\begin{equation}
\label{boundary-charges-comm}
\left[Q, \sigma_{N_0} (\mathbf{x})\right] =
\frac{N_0}{k} \sigma_{N_0}(\mathbf{x}),
\qquad\qquad \left[Q_T, \mu_{M_0} (\mathbf{x})\right] =
\frac{M_0}{k} \mu_{M_0}(\mathbf{x}),
\end{equation}
thus extending earlier results for the Chern-Simons action.
As discussed after \eqref{charge-comm}, these charges are relative
to the area surrounding the point $\mathbf{x}$. 
The reference point $x_0$ in the definition of the fields has also been
omitted according to the previous remarks on the total charge.

%-3.2.1-----------------------------------------------
\subsubsection{3+1d Bulk fields and bulk-boundary correspondence}

The 3+1 dimensional bulk excitations also have associated
charged fields whose form can be derived by extending earlier methods.

\begin{figure}
 \begin{center}
   \includegraphics[scale=.11]{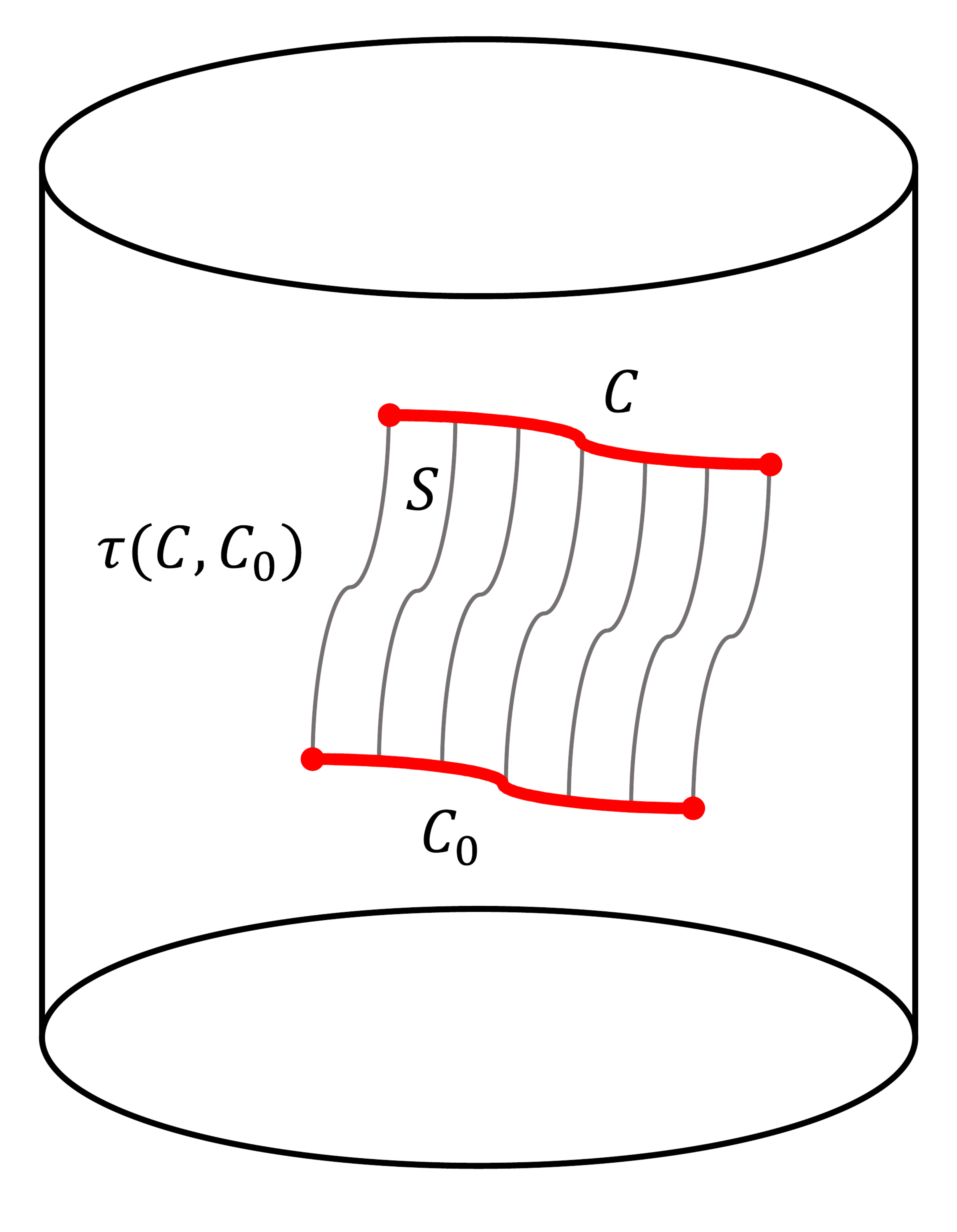}
\includegraphics[scale=.11]{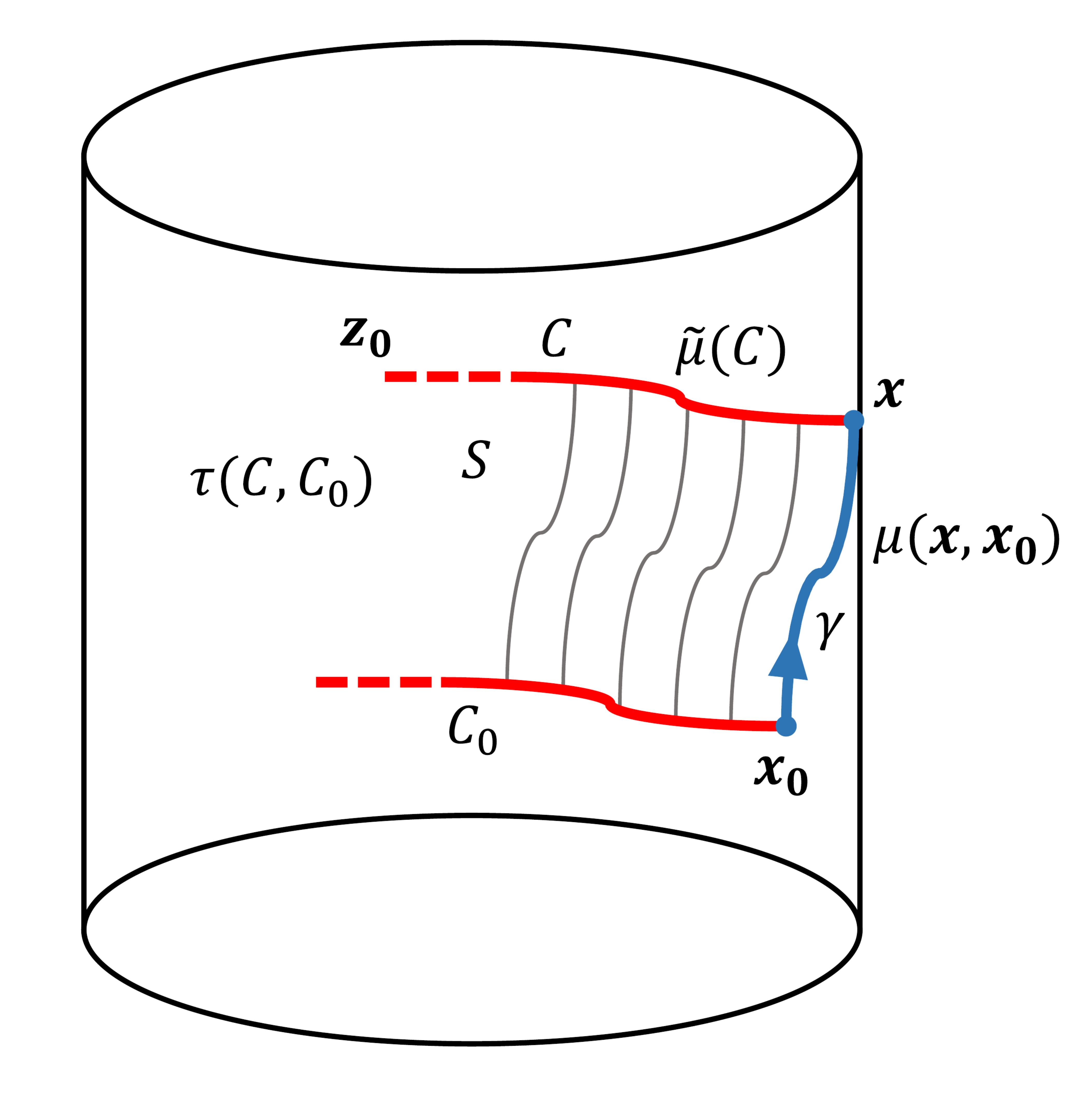}
 \end{center}
 \caption{Topological insulator on the spatial full cylinder. On the
   left: the worldsheet $S$ of the operator $\tau$ \eqref{bulk-sigma-mu},
   representing the time evolution of a vortex line from $C_0$ to
   $C$. On the right: the operator $\tau$ is brought to the boundary,
   where it gives rise to a Wilson line
   $\mu(\mathbf{x}, \mathbf{x}_0)$ with a bulk tail $\wt\mu(C)$
   \eqref{mu-bulk}.}
 \label{fig3}
\end{figure}

We consider the Hamiltonian quantization
of the  bulk BF action (\ref{bulk-action}) for $A=0$.
Its symplectic structure yields the canonical commutators
\begin{equation}
\label{BF-comm-bulk}
\left[ a_i(t, \mathbf{x}), \frac{k}{4 \pi} \epsilon^{jkl}
  b_{kl} (t, \mathbf{y}) \right] = i \delta_i^j
\delta^{(3)} (\mathbf{x}-\mathbf{y}), \qquad\qquad
\Pi^j = \frac{k}{4 \pi} \epsilon^{jkl} b_{kl}\,,
\end{equation}
where now $\mathbf{x},\mathbf{y}$ are three-component spatial vectors.
We introduce open Wilson lines and surfaces operators, as follows,
\begin{equation}
\label{bulk-sigma-mu}
\begin{split}
  & \sigma_{N_0} (\mathbf{x}, \mathbf{x}_0) =\exp
  \left( -iN_0 \int_{\mathbf{x}_0}^\mathbf{x} \d \xi^i a_i (t, \mathbf{\xi})
  \right), \\
  & \t_{M_0} (C, C_0) =\exp \left( -i \frac{M_0}{2} \int_{S, C_0}^C
 \!\!\!   \d \xi^j \wedge \d \xi^\ell \, b_{j\ell} (t, \mathbf{\xi}) \right).
\end{split}
\end{equation}
The field $\sigma$ has been already encountered in the boundary
theory \eqref{sigma-mu-op}, and its dependence on the reference point
$\mathbf{x_0}$ explained.
The operator $\t$ involves the open surface $S$, typically square-like,
representing the `worldsheet' of the vortex line
evolving between boundary lines $C,C_0$, bounded by two further curves
connecting their extrema (see Fig.\ref{fig3}).
Given that $b$ is flat, the shape of $S$ is deformable.
The curve $C_0$ serves as reference for all $\t$ operators,
similarly to $\mathbf{x_0}$ for $\s$, and can be omitted.

The operators \eqref{bulk-sigma-mu} obey the following relations with the bulk 
 charges ${\cal Q}$ and ${\cal Q}_T$ (\ref{bulk-charges})
\begin{equation}
\label{bulk-charges-comm}
\left[{\cal Q}, \sigma_{N_0}\right] = \frac{N_0}{k} \sigma_{N_0},
\qquad\qquad \left[{\cal Q}_T, \t_{M_0} \right] = \frac{M_0}{k} \t_{M_0},
\end{equation}
which are proved  using the commutators \eqref{BF-comm-bulk}.
While $\sigma(\mathbf{x})$ creates the charge $N_0/k$ at point
$\mathbf{x}$, the non-local field $\t$ introduces the vortex line $C$
characterized by the charge density  $M_0/k$ measured by the flux
across the surface $\S$ intersecting $C$ (cf. \eqref{bulk-charges}).

The bulk-boundary relation between the fields $\t$ and $\mu$ in
\eqref{sigma-mu-op}
is now analyzed (see Fig.\ref{fig3}). The vortex worldsheet is assumed to have
one side on the boundary surface, spanning between points $x, x_0$, which
are endpoints for $C, C_0$ on the surface, respectively.
Upon using the bulk equations of motion, $b=\d\Tilde{\z}$, the surface integral
in $\t$ reduces to the line integral on its four boundary lines, two of
which are relevant, \textit{i.e.} not involving reference configurations.
These are the bulk curve $C$ and the surface curve $\g$, spanned by the vortex
endpoint. We thus have the relation (taking $M_0=1$ for simplicity)
\begin{align}
\label{mu-bulk}
\t(C)\sim & \exp \left( -i \int_{\g,\mathbf{x}_0}^\mathbf{x}
    \d \xi^i\, \z_i (t, \mathbf{\xi}) \right)
\exp \left( i \int_{C, \mathbf{z}_0}^\mathbf{x}  \d \eta^i\,
\Tilde{\z}_i (t, \mathbf{\eta}) \right) =
 \mu(\mathbf{x},\mathbf{x}_0) \ \wt\mu(C),
\nonumber\\
\wt\mu(C)= &\exp \left( i \int_{C, \mathbf{z}_0}^\mathbf{x}  \d \eta^i\,
\Tilde{\z}_i (t, \mathbf{\eta}) \right).
\end{align}
Namely, the boundary $\mu$ field \eqref{sigma-mu-op} is obtained,
together with another loop operator
extending from $\mathbf{x}$ into the bulk, corresponding to the vortex line,
expressed in terms of bulk field values $\Tilde{\z}$ (which becomes $\z$ on the boundary).

This bulk `tail' $\wt\mu(C)$ preserves the monodromy phase between $\mu,\s$
when $\s$ is moved slightly in the bulk, and is important for
consistency. Nevertheless, it is a topological term that does not affect
the boundary dynamics of $\mu,\s$ to be discussed in the next section.
The relation \eqref{mu-bulk} between bulk and boundary fields has shown
a difference between the two boundary fields $\mu$ and $\s$.

%-3.3------------------------------------------------
\subsection{Soliton correlation functions and duality}
\label{sec-solitons}

In the previous analysis we found that the $\s,\mu$ fields \eqref{sigma-mu-op}
have the correct charges for soliton configurations.
We now compute their correlators, $\braket{\s(x)\s^\dag(0)}$,
$\braket{\mu(x)\mu^\dag(0)}$ and $\braket{\s(x)\mu(0)}$, 
verify their monodromies, the power-law
behavior of conformal invariance and check the scaling dimensions
with the spectrum of the partition function \eqref{loop-model-Z}. This requires adding the dynamical term \eqref{loop-model-dyn} to the BF action. Notice that this addition does not spoil the symplectic structure \eqref{BF-comm}, because the canonical momentum of the $\z$ field does not change.

The first, direct derivation of the correlators is by inserting the
$a,\z$ Wilson lines in the path-integral with action \eqref{loop-model}
(for $A=0$).
As a preliminary step, we compute the $a,\z$ propagators, by coupling them
to conserved external currents in the action. Upon
evaluating the quadratic path integral, we find
\begin{align}
  \label{pi-int}
  {\bf I}:\  & \braket{a_\mu(x)\z_\nu(y)}=
    i\frac{2\pi}{k}\eps_{\mu\nu\r}\p_\r \left(\frac{1}{-\p^2}\right)_{x,y},
\\
{\bf II}:\ & \braket{a_\mu(x)a_\nu(y)}=0,
  \\
{\bf III}:\ &\braket{\z_\mu(x)\z_\nu(0)}=
          \frac{g_0}{\pi k^2}\frac{\de_{\mu\nu}}{(x-y)^2}.
\end{align}

{\bf I}. In the $\braket{a_\mu\z_\nu}$ expression, one recognizes the kernel
which expresses the linking number of loops $LN(\g,\g')$
\be
\oint_\g dx^\mu\oint_{\g'} dy^\nu \braket{a_\mu(x)\z_\nu(y)}=
\frac{1}{2k}\oint_\g dx^\mu\oint_{\g'} dy^\nu
\eps_{\mu\nu\r}\frac{(x-y)^\r}{|x-y|^3}
=i\frac{2\pi}{k} LN(\g,\g').
\ee
Therefore, the correlator of $\s,\mu$ fields \eqref{sigma-mu-op}, evaluated for closed Wilson loops, is
\begin{equation}
\label{monodromy-sigma-mu}
\braket{\s_{N_0}(\gamma) \mu_{M_0}(\gamma')} =
e^{2\pi i \frac{N_0 M_0}{k}LN(\g,\g')}, \qquad\qquad \p \g =\p \g' =0.
\end{equation}
\medskip
which is the same expression as in the topological theory ($g_0=0$),

{\bf II}. The vanishing result for the correlator $\braket{a_\mu a_\nu}$ is
rather puzzling at first sight, because it would imply
$\braket{\s \s^\dag}_{conn}=0$. Actually, this correlator cannot be computed
directly from the action \eqref{loop-model}, because it is nonperturbative.
If we consider the equivalent action $S[\z]$
(\ref{loop-model-zeta}) in which $\z$ is the only dynamical
field, $a$ is just an auxiliary variable for
writing the solitonic excitation $\sigma = e^{i\int a}$ of $\zeta$.
Namely, $\s$ is a `disorder field' of the theory.

This picture is confirmed by the self-duality of the loop model discussed
in section \ref{sec-duality}, exchanging the fields $a \leftrightarrow \z$
as in (\ref{loop-model-duality}). In a given theory,
say $S[\z]$ (\ref{loop-model-zeta}),
one field is perturbative ($\zeta$) and the other ($a$) is
non--perturbative. Therefore, we shall describe the $\sigma$
correlator by the effects caused on the  configurations
of the $\z$ field: it introduces non-trivial boundary conditions 
in the $\z$ path-integral. Here, we rely on the analysis of disorder
fields in Refs. \cite{frohlich95, frohlich87}.

The field $\s(x_1)$ creates a magnetic flux for the $\zeta$ field at
point $x=x_1$ with local charge
$Q =(1/2\pi)\int_\S \d \z =1/k$, as shown by its commutation relation
\eqref{boundary-charges-comm}.  This flux, in Euclidean covariant
form, is a 2+1d `monopole' with expression
\begin{equation}
\label{monopole}
f_{\mu\nu}(x_1) =\frac{1}{2k} \eps_{\mu \nu \rho}
\frac{(x-x_1)^\rho}{|x-x_1|^3}, \qquad\qquad
   \frac{1}{2\pi}\int_{S^2(x_1)} f(x_1) = \frac{1}{k}\,,
\end{equation}
where $S^{(2)}(x_1)$ is a closed surface surrounding $x_1$ in $\R^3$, obtained by closing the surface $\S\in\R^2$ in the definition of $Q$.
The action $S[\z]$ \eqref{loop-model-zeta} shows the interaction
between such monopoles, which implies that between $\s$ fields.
Therefore, we express the $\braket{\s\s^\dag}$ correlator as the value of the 
Euclidean action in presence of two $\z$ monopoles, as follows\footnote{
  Quantum fluctuations around the classical action
  cancel out in the normalization by $1/Z$. }
\begin{equation}
\label{sigma-corr}
\braket{\s_ {N_0}(x_1) \s_ {N_0}^{\dag} (x_2)} =
\exp\left\{-S[N_0( f(x_1) - f(x_2)]\right\}.
\end{equation}
The evaluation of this action is a rather long computation,
with result (see Appendix \ref{app-A})
\begin{equation}
  \label{sigma-corr2}
  \braket{\s_{N_0}(x_1) \s_{N_0}^\dag(x_2) } = \frac{1}{ (x_1-x_2)^{2\D_\s}},
  \qquad\qquad \D_\s= \frac{1}{2\pi} \frac{N_0^2}{g_0}.
\end{equation}
It has the expected form of conformal field theory, and the
value of the scaling dimension $\D_\s$ matches the spectrum in the
partition function (\ref{loop-model-Z}).  It actually fixes the parameter
$\lambda$ left free in that calculation,
\begin{equation}
    \lambda = \frac{1}{2\pi}.
\end{equation}

\medskip

{\bf III}. The correlator $\braket{\mu \mu^\dag}$ can be calculated
perturbatively from the action $S[\z]$, but this needs further discussion.
A quick result is found by using self--duality: in
analogous way to (\ref{sigma-corr}), the $\mu$ field can be viewed as
generating a monopole of the $a$ field in the dual theory, with
charge $Q_T=(1/2\pi)\int_\S \d a =1/k$ owing to \eqref{boundary-charges-comm}.
This monopole configuration is evaluated with the dual action $\wt S[a]$
(\ref{dual-loop-model}), then $\braket{\mu \mu^\dag}$ is simply found
from (\ref{sigma-corr}) by using the duality of couplings
\eqref{loop-model-duality},
\begin{equation}
\label{mu-corr}
  \braket{\mu_ {M_0} (x_1)\mu^{\dag}_ {M_0}  (x_2)} =
e^{-\wt S[M_0( f(x_1) - f(x_2)]}=  \frac{1}{(x_1-x_2)^{2\D_\mu}},
\qquad \D_\mu=\frac{g_0}{2\pi} \frac{M_0^2}{k^2} .
\end{equation}

We remark that this result can also be understood from
the equations of motion of the action $S[\z]$ \eqref{loop-model}:
\be
\label{loop-model-eq}
 \d a = 0,
\qquad\qquad   i\frac{k}{2 \pi} \epsilon^{\mu \nu \rho}
\p_\nu \zeta_\rho (y) + \frac{g_0}{2\pi} \int \d^3 x\,
    \frac{1}{\p (y,x)} \p_\rho f^{(a)}_{\rho \mu}(x) = 0,
\ee
where $f^{(a)}_{\r\mu}=\p_\r a_\mu -\p_\mu a_\r $.
As seen above, the insertion of $\mu(x_1)$  creates a
monopole for the $a$ field at $x=x_1$, which remains flat outside this point,
$\d a(x)=0, x\neq x_1$. The second equation of motion
shows that the field $\z_\mu(y)$ is never flat for all $y$ values,
due to the nonlocality of the kernel $\p^{-1}$.
It follows that the correlator of two $\mu$ fields is not topological
but depends on the relative distance, leading to the result \eqref{mu-corr}.

The direct, perturbative derivation of $\braket{\mu \mu^\dag}$ is
as follows. The path-integral expression  with two independent
Wilson lines \eqref{sigma-mu-op} can be written,
using the $\braket{\z_\mu \z_\nu}$ kernel \eqref{pi-int}, as
\begin{align}
  \label{mu-pert}
 \braket{\mu_{M_0}(x) \mu_{M_0}^\dag(y)}& =
\left\langle \exp\left(-i M_0 \int^x_{x_0,\g} \z\right)
\exp\left(i M_0 \int^y_{y_0,\g'} \z\right) \right\rangle
  =e^\L,
  \\
  \label{Lambda-int}
  \L &=\frac{g_0 M^2_0}{\pi k^2}\int^x_{x_0,\g} d\xi^\mu\int^y_{y_0,\g'} d\eta_\mu
    \frac{1}{(\xi -\eta)^2}.
\end{align}
The quantity $\L$ strongly depends on the shape of the two lines
$\g,\g'$: on dimensional grounds, it can be estimated
\be
\label{loop-est}
\L\underset{R\gg a}{\sim} -\l\frac{R}{a}+ c \log \left(
  \frac{R}{a}\right) + \dots, \qquad\qquad (\l,c>0),
\ee
where $R,a$ are, respectively, the larger and smaller lengths over
which the integrations in \eqref{Lambda-int} extend, which depend
on the shape of $\g,\g'$ lines.
Actually, once the loop-model interaction is present ($g_0\neq 0$),
these contours can
fluctuate. Thus, the correlator \eqref{mu-pert} should involve a
summation over all possible shapes, while keeping the end-points
fixed. This is to be contrasted with the topological limit ($g_0=0$),
where \eqref{mu-pert} is shape independent and the summation can be
neglected.

We recall that the phase diagram of the loop model is actually
two-dimensional owing to the additional Maxwell term
$t/M\int(\p_\mu\z_\nu-\p_\nu\z_\mu)^2$ \cite{loopmodel,motrunich2012}:
the critical line parameterized by $g_0$ divides two phases
$t\lessgtr 0$ where large loops are suppressed or proliferate.
Therefore, at criticality the linear term in \eqref{loop-est} (perimeter
law) should vanish by a compensation between energy and entropy,
the latter originating from the sum over fluctuating loops
included in \eqref{mu-pert}.
Indeed, the perimeter law is incompatible with conformal invariance.

We evaluate the correlator \eqref{mu-pert} by joining the contours
($ x_0=y_0$) and considering the shortest straight line going between
$x,y$. This configuration is dominating in the phase where large loops
are suppressed, slightly off the critical line.
We then compute \eqref{Lambda-int} for the 
gauge-invariant square of the correlator \eqref{mu-pert} and obtain,
\begin{align}
  \label{mu-corr-2}
 & \langle :\mu_{M_0}(x) \mu_{M_0}^\dag(x+\eps):
   :\mu_{M_0}(y+\eps) \mu_{M_0}^\dag(y): \rangle =
  \left\langle e^{-i M_0 \int^y_{x} \z}\
e^{i M_0 \int^{y+\eps}_{x+\eps} \z} \right\rangle
  =e^\S,
  \\
  & \qquad\qquad \S \underset{|x-y| \gg \eps}{=}
    \frac{g_0 M^2_0}{\pi k^2}\left(\frac{\pi|x-y|}{\eps}
       - \log\frac{(x-y)^2}{\eps^2}\right).
\end{align}
In this calculation, $\eps$ is the distance separating the two 
parallel $(x,y)$ lines, corresponding to the point-splitting of operators.
Assuming that the perimeter term is canceled
by entropy at criticality, we remarkably obtain the
expected conformal correlator \eqref{mu-corr} found earlier
from monopole configurations. That derivation, involving point-like
configurations, luckily did not involve entropy arguments.

In conclusion, we have shown that the loop model dynamics implies
power-law correlators for the soliton fields,
\begin{equation}
  \braket{\s(x) \s^\dag(y)} = \frac{1}{| x-y |^{2\D_\s}},
  \qquad\qquad \braket{\mu(x) \mu^\dag(y)} = \frac{1}{| x-y |^{2\D_\mu}},
\end{equation}
with scaling dimensions $\D_\s,\D_\mu$
in agreement with the expression of the partition function (\ref{loop-model-Z}).

Let us add some remarks.
\begin{itemize}
\item
  {\it Wilson lines as proxies.}
  The simpler derivation of $\mu,\s$ correlators involves the
  monopole configurations \eqref{mu-corr}, \eqref{sigma-corr}, in
  which the Wilson line is only used to create localized charges. This
  is certainly correct in the topological limit of the theory
  ($g_0=0)$: for non-vanishing $g_0$, additional, chargeless terms
  could enter in the definition of the $\mu,\s$ fields to comply with
  conformal invariance. In such a case, Wilson line expressions should
  be simply considered as proxies for the actual conformal primary
  fields.
\item
  {\it Unconstrained Wilson loops as topological defects.}
  We also saw that $\mu,\s$ correlators correspond
  to classes of fluctuating loops having two points $x,y$ fixed, which
  depend logarithmically on $|x-y|$. They should be
  contrasted with closed loops without fixed points, which should describe
  topological defects of the conformal theory. As such, they should not
  depend on the interaction in the loop model action \eqref{loop-model},
  and remain the same as in the topological limit ($g_0=0$).
  A hint of this fact is that monodromy phases are unaffected
  (see \eqref{pi-int}). Furthermore, `hard' defects would
  break conformal invariance. In reality, defects may be topological
  for $g_0\neq 0$ by balancing energy and entropy at criticality.
  We do not have a proof of this fact, but it is natural to assume that
  the conformal theory possesses topological defects.
\item
  Altogether these results let us conclude that the loop
  model is a non--trivial 2+1 dimensional conformal
  field theory. It originates in semiclassical limits of
  electrodynamics, and it displays several interesting nonperturbative
  phenomena in 2+1 dimensions, which can be explicitly analyzed.
\end{itemize}

%-4----------------------------------------------
\section{Fermionic theory on the topological insulator
  boundary}
\label{sec-4}

\subsection{Fermion field}

In this section we introduce the field which represents the fermion in
the loop model. In the following, we shall provide
evidences for this proposal and describe the modifications of the theory which are needed for incorporating fermionic
features. These results will achieve a bosonization of 2+1
dimensional fermions, which holds in the semiclassical approximations
mentioned earlier. This bosonization presents several aspects
in common with the vertex operators construction in 1+1 dimensions.

The fermionic field $\psi(x)$ is made of the solitonic fields
$\sigma_{N_0}$ and $\mu_{M_0}$ \eqref{sigma-mu-op}, which obey
relative statistics with monodromy phase $\Th/2\pi=N_0M_0/k$
\eqref{bulk-phase}.  Hereafter, we assume $k=1$ which identifies the
bulk fermionic response \eqref{response-action}.

The field $\psi(x)$ should be characterized by the following properties.
\begin{itemize}
\item
  Be a product of solitonic fields
  \cite{frohlich95,marino}, so as to have non-trivial statistics, corresponding to half of the monodromy with
  itself $\Th/2\pi=1$.
\item
  Carry unit charge with respect to the $A$ background.
\end{itemize}
We are led to propose the following form
\begin{equation}
\label{fermion-op}
\psi (x) = :\sigma_1(x) \mu_{1/2}(x): =
:\exp\left(-i \int^x_{x_0,\g} \left(a + \frac{\z}{2} \right)\right):\,.
\end{equation}
Let us explain this expression.  The $\sigma(x)$ field adds the charge
$N_0=1$ at $x$, which corresponds to the electromagnetic charge
$Q=1$ for $k=1$ in
(\ref{boundary-charges}). The $\mu_{1/2}(x)$ field has magnetic charge
$M_0=1/2$: it is a kind of `dressing' of $\sigma$ needed
for realizing fermionic statistics. We now compute
the monodromy of the $\psi$ with itself. From the definition
\eqref{fermion-op}, we close the open path $\g$ either
by forming a loop or by taking $x\to\infty,x_0\to -\infty$, thus defining
the field $\psi(\g)$. 
For two fermions with paths $\g,\g'$ we find, using \eqref{monodromy-sigma-mu}
with $N_0=1,M_0=1/2$,
\begin{equation}
\label{fermi-stat}
\braket{\psi(\gamma)\psi(\gamma')} = e^{2\pi i \text{LN}(\gamma, \gamma')}
= e^{i2\pi},
\end{equation}
whose half value is the fermionic statistics.
Note that the two solitons in the definition of a single fermion
$\psi(\g)$ do not braid among themselves
due to the normal-ordering in \eqref{fermion-op}.
Apart from braiding, we assume that $\psi(x)$ does not depend on
the detailed shape of $\g$, according to the discussion of the previous
section. We shall return to this issue in section 4.3.3.

The $\psi$ field \eqref{fermion-op} has scaling dimension
$\D_\s=((1/ g_0+g_0/4)/2\pi)$, using \eqref{sigma-corr2} and
\eqref{mu-corr}: the free-fermion value $\D=1$ is found on the
critical line parameterized by $g_0$. Other values describe
interacting (semiclassical) fermions: indeed, the loop-model conformal
theory involves infinite towers of solitonic states.
The field can also be put in a doublet with its time-reversal partner
$\psi'(x)={\cal T}\psi(x){\cal T}^{-1}=\s(x)\mu_{-1/2}(x)$, since
$\z$ is odd. The action of the time reversal operator ${\cal T}$ on
the doublet $(\psi,\psi')$ and its relation with standard Dirac
spinors will be discussed in the following sections.

We remark that vertex operators of the 1+1d
conformal theory have been similarly derived from 2+1d Wilson lines
in the previous section.
The Weyl fermion was found for $\s=\mu$, while
the Dirac fermion is obtained by a product of $\s(x)$ and $\mu(x)$:
a close inspection shows that the `magnetic' charge should take
half-integer value in this case too.

The definition \eqref{fermion-op} of the fermion operator has one aspect
yet to be understood: the value $M_0=\pm 1/2$ is not present in the
spectrum of the loop model (\ref{loop-model-Z}) which only includes
integer values.
On physical grounds, single electrons should be observable on
the boundary of topological insulators.
We conclude that the bosonic theory studied earlier is
actually missing the states with odd number of fermions,
and thus the quantization condition for the $a$
field should be modified in a sensible way.
Furthermore, a fermionic theory on the geometry $S^2\times S^1$ should
possess two spin sectors which must be identified.

That the BF action (\ref{bulk-action}) is actually not
enough to completely describe a fermionic topological insulator was
already pointed out in \cite{vishwanath2013physics} (see the
introduction and appendix A). Indeed, the problem is that the boundary
theory is actually bosonic and does not (yet) contain fermion excitations. 
The purpose of the rest of this section is to resolve these issues by
applying the recent theory of bosonization, linked to the concept of
higher-form symmetry
\cite{gaiotto2015generalized,gaiottokapustinspinTQFT1,
  gaiottokapustinspinTQFT2}.

%-4.2---------------------------------------
\subsection{Review of modern bosonization}

In this subsection we review the results of Refs.
\cite{gaiottokapustinspinTQFT1,gaiottokapustinspinTQFT2}.  These
studies explained the conditions for the fermionization of a bosonic
theory (and, \textit{viceversa}, for 
 the bosonization of a fermionic
theory).\footnote{This bosonization approach has been developed
  further in \cite{thorngren2020anomalies}, from a more mathematical
  point of view.}
They were meant for topological field theories, but
we shall see that they also apply in presence of loop model dynamics, with
some simplifications which are sufficient for our analysis.

We first recall some basic facts regarding higher-form symmetries
\cite{gaiotto2015generalized},\footnote{ A review with applications to
  condensed matter physics is \cite{mcgreevy2023generalized}. For
  lectures, see \textit{e.g.} \cite{gomes2023introduction} (pedagogical) and
  \cite{bhardwaj2024lectures,luo2023lecture} (advanced). 
  Topological theories of discrete gauge groups are introduced in
  \cite{dijkgraafwitten}.}  which will be used in the following.  We
work in a continuum $d$-dimensional spacetime $\M_d$. A $p$--form
global symmetry with group $G$ is denoted by $G^{(p)}$.  It acts on
extended objects/fields which are $p$-dimensional, like loop ($p=1$)
and surface ($p=2$) operators, by multiplying them by global phases.
Indeed, higher--form symmetries are Abelian
\cite{gaiotto2015generalized}. The conserved current for a $p$--form
symmetry is a $(p+1)$-form, $J_{p+1}$, obeying $d * J_{p+1}=0$. The
corresponding charge $Q({\cal M}_{d-p-1})$ is obtained by integrating
the current on closed sub--manifolds of dimension $d-p-1$, leading to
the following form of the generators of the $p$-form symmetry \be
U_\a({\cal M}_{d-p-1})= \exp\left(i\a Q({\cal M}_{d-p-1})\right),
\quad\quad Q({\cal M}_{d-p-1})=\int_{{\cal M}_{d-p-1}} *J_{p+1}\,,
\label{p-form-def}
\ee
where $\a$ is the group parameter.
The currents $J_{p+1}$ can be coupled to $(p+1)$--form gauge fields
$A_{p+1}$ which are elements of the cohomology group $H^{p+1}(\M;G)$,
rigorously defined on a triangulation of $\M_d$. In the continuum
limit, we can equivalently use flat $U(1)$ $(p+1)$--form fields with
holonomies on $(p+1)$--dimensional closed submanifolds of $\M$ having
values in $G$ \cite{kapustinseiberg2014}:
\begin{equation}
    \d A_{p+1} = 0, \qquad\qquad \exp\left(i \oint A_{p+1}\right) \in G.
\end{equation}
If $G = \Z_N$, the map between continuous and discrete
gauge fields is $A_{cont} = (2 \pi /N) A_{discr}$, where
$A_{discr} \in \Z_N$ takes integer values.

We need the following fact: in $d$ dimensions, gauging a discrete
$p$--form symmetry $G^{(p)}$ leads to the emergence of a dual
$(d-p-2)$--form symmetry $G^{(d-p-2)}$, called dual symmetry or
quantum symmetry.\footnote{Actually, the dual symmetry is
  $\wh{G}^{(d-p-2)}$, with $\wh{G} = Hom (G, U(1))$, the Pontryagin
  dual group. However, for Abelian discrete groups, $\wh{G} \cong G$
  \cite{gaiotto2015generalized,bhardwaj2024lectures}.} Consider
a one--form symmetry $G^{(1)}$, for example, in a theory with
partition function $Z_1$.  This can be coupled to a closed two--form
gauge field $B \in H^2(\M; G)$, leading to $Z_1[B]$. Gauging this
symmetry means taking $B$ dynamical and integrating over it (actually,
summing over it, because $H^2(\M; G)$ is a finite set). So the gauged theory
partition function $Z_2$ is
\begin{equation}
  Z_2 \equiv Z_1^{gauged} = \sum_{B \in H^2(\M; G)} Z_1[B].
  \label{Z-two}
\end{equation}

But then $Z_2$ has a naturally conserved current, $J_{d-2}=*B$, which is
divergentless, since $\d * (*B) = \d B = 0$. This is a $(d-2)$--form current
with conserved
charge defined on codimension $(d-2)$ closed submanifolds
\begin{equation}
  Q(\M_{2}) =  \int_{\M_{2}} B \qquad \ra \qquad U_\a(\M_{2}) =
  e^{i\alpha Q (\M_{2})} \in G^{(d-3)}.
\end{equation}
Therefore, $Q({\cal M}_2)$ generates a $(d-3)$--form symmetry $G^{(d-3)}$,
with $U_\a\in G^{(d-3)}$ being the so-called 'holonomy' of $B\in H^2(\M;G)$.
Thus the claim: $Z_2$ has a global discrete $(d-p-2)$--form symmetry
$G^{(d-p-2)}$, here $p=1$. To conclude this point, we notice that the dual
 of the dual symmetry is the original one:
upon gauging $G^{(d-p-2)}$ we come back to the original
theory with global $G^{(p)}$ symmetry. This is basically due to the
fact that gauging a discrete symmetry is similar to Fourier
transforming the partition function.

We can check this fact in our one--form example: $Z_2$ in \eqref{Z-two}
is coupled to the background gauge field
$C \in H^{d-2}(\M; G)$, via the conserved current $*B$,
\begin{equation}
    e^{i \int_\M *(*B) \wedge C} = e^{i \int_\M B \wedge C}.
\end{equation}
The gauged $Z_2$ theory is then 
\begin{equation}
\label{inv-dual-symmetry}
Z_2^{gauged} = \sum_{C \in H^{d-2}(\M;G)}\,
\sum_{B \in H^2(\M; G)} Z_1[B] e^{i \int_\M B \wedge C} = Z_1\,.
\end{equation}
We used $\sum_C e^{i \int_\M B \wedge C} \sim \delta(B)$, ignoring
normalization factors.  This shows that gauging the dual symmetry
leads back to the original theory. Here the global $G^{(1)}$ symmetry
is generated by
\begin{equation}
  Q(\M_{d-2}) =  \int_{\M_{d-2}} C \qquad \ra \qquad U_\a(\M_{d-2}) =
  e^{i\alpha Q (\M_{d-2})} \in G^{(1)}.
\end{equation}
The fields $C$ and $B$ exchange their roles under this duality.
The argument can be clearly made in reverse.

The second needed fact is a general property of
$(d-2)$--form symmetries in $d$ dimensions. These are generated by
operators $U_\a$ which are integrated on closed $(d-(d-2)-1)$ loci, \textit{i.e.}
one-dimensional loops $W$ (see \eqref{p-form-def}).
These can be interpreted as worldlines of quasiparticles. The loops
$W$ should fuse according to the group law of $G^{(d-2)}$. So, if
a theory possesses quasiparticles whose line operators are
topological and fuse
according to $G$, we can say that the theory has a $G^{(d-2)}$
symmetry. To gauge this symmetry one should sum over all possible
insertions of $W$ operators in the correlation functions, which
means a proliferation of these quasiparticles in the theory.
In physical terms, this is called 'anyon condensation'.

This picture helps to build an intuitive understanding of what is going
on: gauging is equivalent to condensation of the quasiparticles, but
the wavefunction of a condensate is a constant, which has bosonic
statistics. So we can freely gauge a $(d-2)$--form symmetry if the
quasiparticle generators are bosons. If they are fermionic,
the $(d-2)$--form symmetry  can be anomalous and the gauging procedure
ill--defined. This argument is useful for understanding the presence of 
't Hooft anomalies linked to projective representations of the
$(d-2)$--form symmetry $G$ \cite{mcgreevy2023generalized}.

After these remarks, we can now discuss the bosonization in $d$ spacetime dimensions.
We start from a bosonic field theory with partition function $Z_b$
on the manifold $\M_d$.
The fermionic theory partition function $Z_f[\eta]$ also
depends on the choice of spin structure $\eta$ of $\M_d$ (spectrum and
weights are specific for it), while $Z_b$ does not.
Bosonization (fermionization) means writing some
kind of equality $Z_b \sim Z_f[\eta]$.

There are three requirements for this relation to hold:
\begin{itemize}
\item
  in the spectrum of $Z_f[\eta]$ there is the fermion, thus
  in $Z_b$ there should be a quasiparticle which plays this role, with fermionic self--statistics;
\item
  a fermionic theory has always an irreducible $\Z_2$ zero--form
  symmetry, which is the fermion parity $(-1)^F$.
  This should emerge  in the bosonic theory too;
\item
  $Z_f[\eta]$ depends on a spin structure, thus the bosonic theory
  should also implement it.
\end{itemize}

We first focus on the first two. To satisfy the second requirement we
can use the observation made above: if the bosonic theory has a
$(d-2)$--form $\Z_2$ symmetry, by gauging it one gets a global $\Z_2$
zero--form symmetry. But a $(d-2)$--form symmetry is generated by a
quasiparticle, so by parsimony we can just assume that precisely the
fermionic quasiparticle generates this $\Z_2^{(d-2)}$ symmetry. We
conclude that in the bosonic theory there should be a quasiparticle
with fermionic self--statistic which fuses with itself into the
identity ($\Z_2$). As a consequence though, this $\Z_2^{(d-2)}$
symmetry is actually anomalous and cannot be gauged.\footnote{In
  \cite{gaiottokapustinspinTQFT1}, the argument is actually made in
  reverse: by explicit construction of a large class of spin--TQFTs,
  the Authors show how the bosonic TQFT should have this specific
  anomaly. Then, they notice that this pertains to 
  the symmetry of a fermionic quasiparticle which self--fuses into the
  identity.} We remark that this anomaly is independent from the
specific theory considered\footnote{Let us consider two different
  theories QFT1 and QFT2 with fermionic quasiparticles $\psi_1$,
  $\psi_2$. If they are decoupled, the product $\psi_1\psi_2$ is a
  bosonic particle which generates a $\Z_2^{(d-2)}$ symmetry, non
  anomalous. Thus the two anomalies are the same:
  they are two equal signs that square to one.}
and it has a precise form, given by the Steenrod
square $Sq^2(B)$ of the background $(d-1)$-gauge field $B$
\cite{gaiottokapustinspinTQFT1}. This fact implies that the anomaly occurs for $d>2$, while in $d=2$ the
$\Z_2^{(0)}$ symmetry is not anomalous: we can argue that, in
the condensation picture, there is no real difference between
fermions and bosons. Finally, the would--be anomaly
is trivialized by the introduction of a spin structure (the last of
the three points above).

All in all, we are led to the following conclusion. A
bosonic theory can be fermionized if it possesses a
$\Z_2^{(d-2)}$ anomalous global symmetry with a specific form for the
anomaly. Then, gauging this symmetry yields the fermionic theory, by
writing the following relation between the corresponding partition functions
\begin{equation}
\label{general-bosonization}
    Z_f [\eta]= \sum_{B \in H^{d-1}(\M; \Z_2)} Z_b[B] z[B,\eta].
\end{equation}
The extra factor $z[B,\eta]$ has two features: it has the same anomaly
as $Z_b$, in order to cancel it and allow for the gauging of
$\Z_2^{(d-2)}$, and it introduces a spin structure $\eta$. In
\cite{gaiottokapustinspinTQFT1}, the general form for $z[B,\z]$ is
given. This is the general procedure used in
\cite{gaiottokapustinspinTQFT2} to construct a spin--TQFT out of the
bosonic TQFT, called 'shadow theory'.

The relation (\ref{general-bosonization}) can be inverted by gauging
the fermion parity $\Z_2^{(0)}$ in $Z_f$, as in (\ref{inv-dual-symmetry}).
This means coupling $Z_f$ to a dynamical gauge field
$s \in H^{1}(\M; \Z_2)$, which is an assignment of holonomies on
non--trivial loops in $\M$: $e^{i \oint s} \in \{1,-1\}$.  It amounts
to a choice of periodic or antiperiodic boundary conditions for
fermions on non--trivial loops, \textit{i.e.} a choice of a spin
structure $\eta$ on $\M$. Thus gauging fermion parity is equivalent to
sum over all spin structures (indeed, the number of inequivalent spin
structures on a spin manifold $\M$ is $|H^1(\M, \Z_2)|$). Therefore,
starting from a fermionic theory, one obtains a bosonic one by summing
over all spin structures, a well--known result
of bosonization in 1+1d conformal theory \cite{ginsparg1988applied}.
 
We can relate what has been said in this subsection with the duality
maps between fermionic and bosonic theories in 2+1 and 1+1 dimensions
\cite{wittenwebofduality,tongwebofduality,tongArf}.  In 2+1d, the
standard way to obtain fermions in a bosonic theory is by flux
attachment, \textit{i.e.} coupling the theory to a Chern--Simons
term. Indeed, the Chern--Simons action introduces a spin structure
dependence in the bosonic theory and it is a common tool to generate
anomalies in odd dimensions \cite{wittenwebofduality,
  tongwebofduality, kapustinExactbosonization}. In 1+1d, the term
$z[B,\eta]$ is the Arf invariant \cite{gaiottokapustinspinTQFT1},
which is the basic quantity used to realize 1+1d dualities in
\cite{tongArf}. 
The Arf invariant is also linked to the flux
attachment in one dimension higher, being related to the Chern--Simons
action by dimensional reduction (see Appendix C of \cite{tongArf}).

%-4.2.1--------------------------------------------

\subsubsection{The toric code and its anomaly}

The toric code provides an interesting example of bosonization/fermionization
in 2+1 dimensions which has been considered in the literature \cite{gaiottokapustinspinTQFT1,gaiottokapustinspinTQFT2}.
We are interested in determining the form of the anomaly for the
$\Z_2^{(1)}$ symmetry discussed before.
As said, the bosonic theory should possess a fermionic quasiparticle
which generates this symmetry. The toric code is a lattice $\Z_2$
gauge theory, whose low--energy behavior is described by a $k=2$ BF
topological theory, neglecting all dynamical aspects of the theory.
There are two quasiparticles excitations $e$, $m$,
whose composition gives the fermion we are looking for.

The BF theory is, 
\begin{equation}
  S = i\frac{k}{2\pi} \int_{{\cal M}_3} \zeta \d a,
  \label{BF-again}
\end{equation}
allowing temporarily general $k$ values and choosing
$\M_3=S^1\times S^1\times \R$, for example.
The electric and magnetic quasiparticles, respectively $e,m$,
are associated to the Wilson loops,
\be
W[a]_i=\exp\left(i\int_{\g_i} a \right), \qquad\qquad
W[\z]_i=\exp\left(i\int_{\g_i} \z \right),
\ee
where the $\g_i$, $i=1,2$, are non-trivial cycles of ${\cal M}_3$.

We recall that the Hamiltonian quantization of the BF theory on the
spatial torus determines the following algebra 
\be
W[a]_iW[\z]_jW[a]_i^{-1} =\exp\left(i\frac{2\pi}{k}\right) W[\z]_j,
\qquad i\neq j=1,2,
\label{clock}
\ee
whose representation has minimal dimension $k^2$ on this geometry.

The BF theory possesses two global one-form symmetries defined by
shifting the fields $\zeta \ra \zeta + \lambda$ and $a \ra a + \alpha$
by closed forms $\lambda$, $\alpha$.  The action \eqref{BF-again} is
indeed invariant (the fields $a,\z$ themselves being closed due to the
equations of motion).  The loops $W[a],W[\z]$ are multiplied by global
phases under these symmetries.

The algebra \eqref{clock} shows that the symmetry
group is $\Z_k^{(1)} \times Z_k^{(1)}$. One kind
of loops, say $W[a]_i$, are generators of the $\Z_k^{(1)}$ acting
on the other ones, $W[\z]_j$, which are charged with respect to this symmetry,
thus realizing the pattern of higher-form symmetries
discussed in the previous section.
The two types of loops exchange their role for the other $\Z_k^{(1)}$ symmetry.

We now remark that each $\Z_k^{(1)}$ symmetry is non-anomalous and can be
independently gauged. This is obtained by coupling one current, say
$J=* a$, to a background two--form gauge field $B$, which is a flat
and obeys $e^{i\oint B}\in \Z_k$, as it should.  The gauged
action is 
\begin{equation}
  \label{single-sym}
    S = i\frac{k}{2\pi} \int_{\M_3}  \zeta \d a + a B\,,
\end{equation}
and is left invariant by the transformation $\z\to\z-\b$ and $B\to B+d\b$,
where $\b$ now is a generic one-form.

Next we try to gauge the diagonal $\Z_k^{(1)}$ symmetry
which is generated by the product of the two kinds of loops. The action is now
\begin{equation}
  \label{diag-symm}
    S = i\frac{k}{2\pi} \int_{\M_3}  \zeta \d a + (\zeta + a) B\,,
\end{equation}
and the gauge transformations are
\begin{equation}
  \label{Z2-gauge}
  \zeta \ra \zeta - \beta, \qquad  a \ra a -\beta, \qquad  B \ra B + \d \beta. 
\end{equation}
These do not leave the action invariant,
\begin{equation}
  S\to S+\D S, \qquad   \D S = - i\frac{k}{2\pi}
  \int_{{\M}_3} 2\beta B + \beta \d \beta.
 \label{toric-code-anomaly}
\end{equation}
Therefore, we have found the mixed ('t Hooft) anomaly which prevents gauging,
in agreement with the general discussion of the previous section.

We now specialize to the $k=2$ case of the toric code. The fermion
is identified with the product of the two quasiparticles,
$\psi=em=\exp(i\oint a +\z)$, and the anomaly \eqref{toric-code-anomaly}
should be canceled in the gauged partition function
\eqref{general-bosonization}.

Actually, this is possible by adding a term to the bulk topological
action, a rather standard fact in our setting of topological
insulators.  We assume that the flat $B$ field extends in 3+1
dimensions and write the action
\begin{equation}
\label{anomaly-action}
S_{bulk} =i \frac{2}{2\pi} \int_{\M_4} B \wedge B, \qquad\qquad
B \in Z^2(\M_4; \Z_2)\,.
\end{equation}
This expression, with $\M _3 = \p \M_4$, reproduces the anomaly
(\ref{toric-code-anomaly}) under
$B \ra B + \d \beta$. It is instead gauge invariant if
$\p \M_4 = \emptyset$.

The quantity \eqref{anomaly-action} is the Steenrod square $Sq^2(B)$
we were alluding to before (in a special case).
It only takes values $0, \pm i\pi$, thus
signs are not important in its expressions.
It depends on the spin structure of $\M_d$ (here $d=2+1$),
thus we should require that the extension to 
$\M_{d+1}$ is consistent, namely that this is an orientable manifold
to which the spin structure of $\M_d$ extends
\cite{gaiottokapustinspinTQFT1,kapustinthorngren2017}.
Under these conditions, the expression \eqref{anomaly-action} is
well defined. Actually, it is a three-dimensional quantity, although written
in four dimensions.

\bigskip

{\bf On continuum and discrete formulations.}
The $\Z^{(1)}_2$ anomalies \eqref{toric-code-anomaly} and
\eqref{anomaly-action}
have been found by using differential calculus in the continuum,
assuming that holonomies of flat forms take values in discrete groups.
This formulation is useful for the later discussion of the loop model and
its dynamical aspects.
A more rigorous descriptions of the anomalies is discrete: 
a triangulated topological space $\M$ is considered, on which a discrete $G$
background field is an element $[B]$ of the cohomology group
$H^{p+1}(\M; G)$ \cite{gaiottokapustinspinTQFT1}. In actual
computations, one chooses a representative $B$ of the cohomology class
$[B]$, which is a cocycle $\delta B=0$, $B \in Z^2(\M;G)$.
In this setting, one defines the
coboundary operator $\delta$ ($\delta^2=0$) and the cup product
$\cup$, which respectively replace the differential $\d$ and the wedge product
$\wedge$ of differential calculus in the continuum.

Our correspondence between discrete and continuum formulations is \textit{i.e.},
for $\Z_2$,
\begin{equation}
  \delta \ra \d, \qquad \cup \ra \wedge, \qquad
  B_{\text{continuum}} = \frac{2\pi}{2}B_{\text{discrete}} =
  0, \pi, \qquad {\rm mod}\ 2\pi.
\end{equation}
However, the map is nontrivial: for example,
the cup product is not graded commutative at the cochain level (contrary to the wedge product), and there
are higher cup products which do not have clear counterparts in
differential calculus.
In the following, we rederive the anomalies in discrete form
for showing the differences, and  eventually argue
that both formulations are equivalent for the sake of our analysis.

The discrete formulation of the BF theory \eqref{diag-symm} is $(k=2)$
\be
\label{diag-discr}
S=i\pi\int \z \cup\de a +B\cup a +\z\cup B \,.
\ee
The terms are witten paying attention to the non-commutativity of $\cup$,
in order to unsure gauge invariance for each $\Z^{(1)}_2$, as in
\eqref{single-sym}.
It follows that the anomaly for the diagonal subgroup is
\cite{gaiottokapustinspinTQFT1}, 
\be
\label{cup-anomaly}
\D S = i\pi \int B \cup \beta + \beta \cup B +  \beta \cup \delta \beta.
\ee
The compensating bulk action is, under gauge transformations
$B \to B - d\b$\,.
\be
\label{anomaly-action2}
S_{bulk}= i\pi \int B \cup B.
\ee

We remark that the first two terms in the anomaly \eqref{cup-anomaly}
do not sum up as in \eqref{toric-code-anomaly}: reversing the ordering
would require the following identity, valid
for $A$, $B$ general cochains,\footnote{
  Up to irrelevant signs for $\Z_2$ valued objects.}
\begin{equation}
\label{cup-product}
B \cup A - A \cup B = \delta(A \cup_1 B) +
\delta A  \cup_1 B + A \cup_1 \delta B, 
\end{equation}
which introduces the higher cup product $\cup_1$
\cite{gaiottokapustinspinTQFT1}.  Nonetheless, the continuum and
discrete descriptions work in parallel to show the cancellation of the
anomaly by the bulk action, which takes the equivalent forms
\eqref{anomaly-action} and \eqref{anomaly-action2}.  In conclusion, we
find that the continuum formulation of discrete symmetries is
consistent for our purposes.

%-4.3----------------------------------------------
\subsection{Fermionic theory from the loop model}

In this section we apply the bosonization procedure presented in
section 4.2 to the loop-model theory on the boundary of topological
insulators.
We remind the main steps involved:
\begin{itemize}
\item To identify the $\Z_2^{(1)}$ symmetry generated by
fermionic excitations and its anomaly.
\item
  To gauge this symmetry and obtain the fermionic partition function
with $Z_2^{(0)}$ fermion-number  symmetry and spin sectors.
\end{itemize}

\subsubsection{$\Z_2^{(1)}$ symmetry and its anomaly}
The first step will take advantage of previous results for the toric
code. The loop model presents two new features with respect to this
theory: in the action \eqref{loop-model}, the BF term  is
accompanied with a dynamical term not present before; furthermore,
the fermionic theory should correspond to $k=1$ (we shall return to
the $k=2$ case of the toric code in section 4.6).

We remark that the dynamical term in the loop model action \eqref{loop-model}
does not modify the global symmetries of the  BF
part. Indeed, the equations of motion (\ref{loop-model-eq}) still imply
$\d a = \d \zeta = 0$ in the vacuum and the
monodromy between closed Wilson loops (\ref{monodromy-sigma-mu})
is given by the BF part of the action.
Furthermore, the action is invariant under
$\zeta \ra \zeta + \lambda_1$ and $a \ra a + \lambda_2$, with
$\lambda_{1,2}$ closed one--forms. Thus, the loop model for coupling
$k$ also possesses the $\Z_k^{(1)} \times \Z_k^{(1)}$ symmetries.

The fact that we consider the $k=1$ case may seem puzzling, because there are
apparently no one--form symmetries. However,
we should find the symmetry generated by the fermion,
which is associated to a half-integer charge,
$(N_0,M_0)=(1,1/2)$. The fermion closed loop
$\psi(\g)=\exp(-i\oint_\g a +\z/2)$ possesses fermionic self--statistics
and actually generates a $\Z_2^{(1)}$ symmetry.
The associated current for this global symmetry is
\begin{equation}
    J=*\left(a + \frac{1}{2}\zeta\right),
\end{equation}
and the $k=1$ loop-model action (\ref{loop-model}) is invariant under
\begin{equation}
\label{global-Z2-symm}
a + \frac{1}{2}\zeta \ra a + \frac{1}{2}\zeta - 2\l , \qquad
\Rightarrow \qquad a \ra a - \l, \quad \zeta \ra \zeta -2\l,    
\end{equation}
with $\l$ a closed one--form.

We now proceed to gauge this symmetry by coupling the theory to the
$\Z_2$ flat two-form field $B$, as in the case of the toric code.
The loop model action \eqref{loop-model} (taking $A=0$ momentarily)
is modified into
\begin{equation}
\label{loop-model-B}
S =\frac{i}{2\pi}\int_{\M_3}  \zeta \d a +
2\left(a +\frac{1}{2}\zeta \right) B+
\frac{g_0}{8\pi}\int_{\M_3} \left(f_{\mu\nu}+B_{\mu\nu} \right)
\frac{1}{\partial}  \left(f_{\mu\nu}+B_{\mu \nu}\right)\,,
\end{equation}
where $f = \d a$. Under the local version of the transformation
(\ref{global-Z2-symm}),
\begin{equation}
\label{gauged-Z12-symm}
a \ra a - \beta, \qquad \zeta \ra \zeta -2\beta,
\qquad B \ra B + \d \beta,
\end{equation}
the dynamical term of the action (\ref{loop-model-B}) is gauge invariant,
while the topological part changes as follows 
\begin{equation}
\label{anomaly-B}
S \ra S + \D S, \qquad  \D  S = -\frac{i}{\pi}
\int_{\M_3} 2\beta B + \beta \d \beta \,.
\end{equation}
This term is precisely the mixed anomaly
(\ref{toric-code-anomaly}) found in the toric code, which
can be canceled by the $B^2$ bulk term \eqref{anomaly-action}.
The first step of the bosonization program is thus achieved.

We recall that the field $B$ realizes $\Z_2$ holonomies,
implying the quantization
\begin{equation}
\label{B-val}
\frac{1}{2\pi}\int_{S^2} B=0, \frac{1}{2}, \quad  \text{mod } 1 , \qquad
\Rightarrow \qquad e^{i\int_{S^2} B } \in \Z_2\,,
\end{equation}
in the spatial $S^2$ geometry we have been considering.  Let us see
how this modifies the original solitonic charge assignments of the loop model
\eqref{boundary-charges}. The equations of motion are
\begin{equation}
\label{loop-model-B-eqs}
    B = -\d a, \qquad\qquad B = -\frac{1}{2}\d \zeta.
\end{equation}
These equation should not be considered as local conditions, because
$B$ is not an Abelian field, but a continuous representative of a $\Z_2$
sign. Therefore, they only impose a parity constraint on the
$a,\z$ fluxes as follows
\begin{equation}
\label{N0-B}
-\frac{1}{2\pi}\int_{S^2} B = \frac{1}{4\pi} \int_{S^2}\d \zeta =
\frac{N_0}{2} =0, \frac{1}{2}, \quad  \text{mod } 1,
\end{equation}
\begin{equation}
\label{M0-B}
-\frac{1}{2\pi}\int_{S^2} B = \frac{1}{2\pi} \int_{S^2}\d a =
M_0 =0, \frac{1}{2}, \quad  \text{mod } 1.
\end{equation}

More precisely, the theory (\ref{loop-model-B}) is characterized by
two sectors in the geometry $S^2 \times S^1$, which have the following
$N_0$, $M_0$ quantizations, according to the periods of $B$ in
(\ref{N0-B}), (\ref{M0-B}):
\begin{itemize}
    \item the `even' sector
    \begin{equation}
    \label{even-sector}
    \frac{1}{2\pi}\int_{S^2} B = 0, \qquad \to \qquad
    N_0 =2n ,\quad M_0 =m, \qquad n,m\in \Z\,;
    \end{equation}
    \item the `odd' sector
    \begin{equation}
    \label{odd-sector}
    \frac{1}{2\pi}\int_{S^2} B = \frac{1}{2}, \qquad \to \qquad
    N_0 =2n+1 ,\quad M_0 =m+\frac{1}{2}.\qquad\quad
    \end{equation}
\end{itemize}
We  have found  that the  presence  of the  $B$ field  allows for  the
half-integer  quantization of  $M_0$, which  was not  possible in  the
original  theory  \eqref{loop-model},  but  is  indeed  necessary  for
defining  the  fermion  field   (\ref{fermion-op}). Actually, the even
and odd sectors contain corresponding numbers of  fermions.

The $B$ field does not lead to other effects in the loop model, 
 as shown by the following argument.
The electromagnetic
field $A$ can be introduced in the modified action \eqref{loop-model-B}
by respecting the $\Z^{(1)}_2$ gauge invariance (\ref{gauged-Z12-symm}),
namely by coupling it to the gauge invariant combinations
\begin{equation}
\label{a-zeta-tilde}
   \d a + B=  \d \tilde{a}, \qquad\qquad \d \zeta + 2B =\d \Tilde{\zeta} ,
 \end{equation}
 which have been rewritten in terms of tilded variables owing to the
 flatness of $B$.
The full expression for the loop model action is therefore
\begin{equation}
\label{loop-action-full}
\begin{split}
  S =&  \frac{i}{2\pi}\int_{\M_3}  \zeta \d a +
 2\left(a +\frac{1}{2}\zeta \right) B +
 \left(\d \zeta +2B \right) A - \frac{1}{4}(\d a + B)A 
  \\
  & + \frac{g_0}{8\pi}\int_{\M_3}  (f +B)_{\mu\nu}
  \frac{1}{\partial}  (f+B)^{\mu\nu} ,
\end{split}
\end{equation}
where $f_{\mu\nu}= \p_\mu a_\nu -\p_\nu a_\mu$. It 
also includes the parity anomaly term, and should 
be compared with \eqref{top-surface-action}, \eqref{loop-model}
for $k=1$.
Using the tilded variables \eqref{a-zeta-tilde}, we can rewrite this action as 
\begin{equation}
\label{tilde-act}
S =
 \frac{i}{2\pi}\int_{\M_3}  \Tilde{\zeta} \d \Tilde{a} +
 \Tilde{\zeta} \d A  - \frac{1}{4} \Tilde{a} \d A 
 + \frac{g_0}{8\pi} \int_{\M_3} \Tilde f_{\mu\nu}
  \frac{1}{\partial}  \Tilde f^{\mu\nu} 
 +  \frac{i}{\pi}\int_{\M_4} B^2 \,.
\end{equation}
This expression is rather interesting: it has the
same form as the original one (\ref{loop-model}) in absence of the $B$
field, plus the extra bulk term $B^2$ which originates from the BF action
rewritten as a total derivative in one extra dimension, actually the bulk,
$\p\M_4=\M_3$.

We can also integrate the
dynamical fields $\zeta$ and $a$ (but not $B$), to obtain
the  boundary response action
\begin{equation}
\label{response-full}
S_{ind}[A,B] = \frac{g_0}{8 \pi} \int_{\M_3} F_{\mu\nu}
\frac{1}{\p} F^{\mu\nu} - \frac{i}{8\pi} \int_{\M_3} A \d A
+ \frac{i}{\pi} \int_{\M_4}  B^2.
\end{equation}
This is again the one--loop fermionic response (\ref{loop-ind-action}),
plus the parity-anomaly term and the $\Z^{(1)}_2$ anomaly.
Both are canceled by the topological bulk theory, as better discussed
later.

To conclude, it is apparent that the $B$ field has been hidden
in the new connections (\ref{a-zeta-tilde}), and its only effect is
to modify the Dirac quantization conditions 
(\ref{boundary-charges}) into \eqref{N0-B}, \eqref{M0-B}.
We remark that the map to tilded variables $(a,\z)\to(\Tilde a,\Tilde \z)$
introduced by \eqref{a-zeta-tilde} is well defined because $B$ is a flat
connection. Global issues of $B$ will be analyzed in the following.

%-4.3.2--------------------------------------------------

\subsubsection{Gauging the $\Z_2^{(1)}$ symmetry}
The second step of bosonization is to transform the $B$ field
from a background to a dynamical degree of freedom, \textit{i.e.}
to gauge the $\Z_2^{(1)}$ symmetry.
This is achieved by summing the bosonic partition function
over the two $B$ sectors,\footnote{
  As said before, we can regard (\ref{gauged-Z}) as a three--dimensional theory
  \cite{gaiottokapustinspinTQFT1,kapustinthorngren2017}. } as outlined in \eqref{general-bosonization},
\begin{equation}
\label{gauged-Z}
Z_f[\eta] = \sum_{B \in H^2(S^2\times S^1; \Z_2)} Z[B]
\exp\left(\frac{i}{\pi}\int_{\M_4} B^2 \right).
\end{equation}

The phase factor cancel the $\Z_2^{(1)}$ anomaly of the theory
\eqref{response-full}, thus allowing the correct definition of
$Z_f[\eta]$. The two spin structures on $S^2 \times S^1$ are given by the
choice of periodic or anti--periodic boundary conditions for fermions
around the time $S^1$: we can call $\eta_+$ the natural anti--periodic
conditions, \textit{i.e.}  $Z_f[\eta_+] = Z_f^{NS}$, and $\eta_-$ the periodic
conditions, $Z_f[\eta_-]=Z_f^{\widetilde{NS}}$. 

We recall from Section 4.2 that gauging a discrete symmetry is an 
invertible process.
The $Z^{(0)}_2$ fermion-number symmetry can be gauged back
by introducing a one-form  $\Z_2$ field $s \in H^1(\M_3; \Z_2)$,
having holonomies around
the time circle $\int_{S^1} s = \{0, \pi \}$.
It follows that introducing $s$ as a background, and turning it
one can pass from one spin structure to the other:
it effectively changes the boundary conditions,
from $\eta_+$ to $\eta_-$ and
\textit{viceversa}. So the two possible fermionic partition functions
are
\begin{equation}
\label{Z_f}
Z_f[\eta_+]= Z[\eta; {\textstyle\int_{S^1}}s =0] = Z_f^{NS}, \qquad\quad
Z_f[\eta_-]= Z[\eta, {\textstyle\int_{S^1}}s = \pi] = Z_f^{\widetilde{NS}}.
\end{equation}

The background field $s$ can be added by coupling it to the
conserved current for the fermionic parity $J=*B$.
In conclusion, a more explicit form of the fermionic partition function
is given by
\begin{equation}
\label{Z_f-S}
Z_f[\eta,s] = \sum_{B \in H^2(S^2\times S^1; \Z_2)} Z[B]
e^{\frac{i}{\pi} \int_{\M_4} B^2 +\frac{i}{\pi} \int_{\M_3} B s}.
\end{equation}
In this expression, considering the two possible values of $s$
leads to the partition functions (\ref{Z_f}).
Let us show their explicit expressions.

The bosonic partitions $Z[B]$ have the form (\ref{loop-model-Z}),
where $B$ only affects the quantization
conditions \eqref{even-sector}, \eqref{odd-sector}, as discussed before.
We have
\begin{equation}
\label{Z_B}
\begin{split}
  &Z[{\textstyle\int} B=0] = Z_{osc}\sum_{(N_0,M_0) = (2n,m)}
  e^{-\frac{\beta}{2 \pi R} \left(\frac{N_0^2}{g_0} + g_0 M_0^2 \right)},
  \\
  &Z[{\textstyle\int} B=\pi] =Z_{osc} \sum_{(N_0,M_0) = (2n +1,m+1/2)}
  e^{-\frac{\beta}{2 \pi R} \left(\frac{N_0^2}{g_0} + g_0 M_0^2 \right)},
\end{split}
\end{equation}
where $n,m \in \Z$.

The evaluation of the phase factor in \eqref{Z_f-S} is as follows.
The $B^2$ term cancels with the anomaly.
The coupling $i/\pi\int Bs$ is non zero when both
fields are non--trivial:
\begin{equation}
  \frac{1}{\pi}\int_{S^2 \times S^1} B s =
  \frac{1}{\pi}\int_{S^2} B \int_{S^1} s =
    \begin{cases}
        \pi \quad \text{if } \int_{S^2} B = \int_{S^1} s = \pi;\\
        0 \quad \text{otherwise}.
    \end{cases}
\end{equation}
This means that when $\int s=\pi$, the odd sector in
(\ref{Z_B}) gets a minus sign in the sum for the partition
function.

We thus find
\begin{equation}
\label{Z_f-NS}
    Z_f^{NS} =  Z[0] + Z[\pi], 
\end{equation}
\begin{equation}
\label{Z_f-NStilde}
Z_f^{\widetilde{NS}} =  Z[0] - Z[\pi] .
\end{equation}
In explicit form, we have
\begin{equation}
\label{Z_f-k2}
Z_f[\eta_\pm] = Z_{osc}\left( \sum_{(n_0,m_0) \in 2\Z}
  \pm \sum_{(n_0,m_0) \in 2\Z+1}\right)
\exp\left({-\frac{\beta}{2 \pi R} \left(\frac{n_0^2}{g_0} +
      g_0 \frac{m_0^2}{4} \right)}\right),
\end{equation}
where we introduced the integer labeling,  
$N_0 = n_0$, $M_0 = m_0/2$, $n_0, m_0\in \Z$, and 
$\eta_\pm$ denote the spin structures of $NS$ and $\wt{NS}$ sectors.

This expression $Z_f[\eta_\pm]$ is an important result of this paper, because it
clearly identifies the 2+1d fermionic conformal theory on the
boundary of topological insulators. The partition function and the effective action
\eqref{loop-action-full} 
correctly account for the topological and geometrical aspects of fermions.
For what concern the dynamics, $Z_f[\eta_\pm]$ is limited by the semiclassical
nature of the loop model: for example, there are infinite
towers of stable solitons, and fluctuations in $Z_{osc}$ \eqref{Z-osc}
have multiplicities of Bose statistics.
 
The expressions \eqref{Z_f-NS}, \eqref{Z_f-NStilde} can  be inverted to obtain 
\begin{equation}
\label{Z_b-sum}
Z[0] = \frac{1}{2}(   Z_f^{NS} + Z_f^{\widetilde{NS}}),
\qquad\quad Z[\pi] =  \frac{1}{2}( Z_f^{NS} - Z_f^{\widetilde{NS}}).
\end{equation}
The quantity $Z[0]$ realizes the sum over spin structures, corresponding to
integrating (\ref{Z_f-S}) 
over $s$ (\textit{i.e.} gauging the fermionic parity
by making $s$ a dynamical gauge field). This gives back the
original ungauged theory with $B =0$, see (\ref{inv-dual-symmetry}),
\begin{equation}
\label{Z_B0}
\frac{1}{2}\sum_{ \eta} Z_f[\eta] \equiv\frac{1}{2}
\sum_{s \in H^1(S^2 \times S^1; \Z_2)} Z_f[\eta, s] =  Z[0],
\end{equation}
which is the first expression in \eqref{Z_b-sum}. In a Hamiltonian
formulation, it can be written as
\begin{equation}
  \label{Z-sum-sect}
    Z[0] =   \frac{1}{2} \text{Tr}\left[ (1 + (-1)^F)e^{-\beta H} \right],
\end{equation}
where the trace is taken over the fermionic Hilbert space on the spatial
$S^2$. The factor $(-1)^F$ occurs for periodic boundary conditions
in the time direction $S^1$. The partition function (\ref{Z_B0})
only involves states with an even number of fermions, which
are then bosonic.

The sum over spin structures also occurs in two--dimensional
bosonization, and is required by modular invariance of the
partition function on the torus $S^1\times S^1$ 
\cite{ginsparg1988applied}.  This is the standard setting for
applications to statistical mechanics, unless special boundary
conditions are considered. 
In the quantum Hall effect and other
topological phases of matter, single fermion excitations should be
present: in these cases, the requirement of modular invariance should
be relaxed \cite{cappelli1997}.  Note also that the bosonic description of 1+1d Dirac
fermions for each independent spin sector has been obtained only
recently within the general approach described here \cite{tongArf}.

We remark that the original bosonic spectrum
\eqref{loop-model-Z}  also shows integer-valued excitations, but is 
different from that of $Z[0]$ \eqref{Z-sum-sect}, which is constrained by
the  condition on $\s,\mu$ forming fermion composites. 
In the odd sector $Z[\pi]$ in \eqref{Z_B}, the charge $M_0$  
takes half--integer values.
This possibility was not considered in the earlier analysis
\cite{loopmodel}, neither in other works on fermionic
topological insulators \cite{zirnstein2013cancellation,
  chen2016bulk}. In \cite{ye2015vortex}, the authors consider standard
integer quantization in the description of bosonic
topological insulators.

%-4.3.3---------------------------------------

\subsubsection{Gauge invariance of fermionic excitations}

The definition of the fermion operator \eqref{fermion-op} should be
reconsidered after the introduction of the $B$ field and the
gauging of the $Z^{(1)}_2$ symmetry discussed in the previous sections.
The gauge invariant fields are now $\Tilde{a}, \Tilde{\z}$,
with $d\Tilde{\zeta}=\d \zeta + 2 B$ and
$d\Tilde{a}=\d a + B$, and Wilson loop operators should be written in terms of
them. The fermion now reads, for closed loops,
\begin{equation}
\label{fermion}
\psi[\gamma] = e^{-i \oint_\gamma \Tilde{a} + \frac{1}{2}\Tilde{\zeta}} =
e^{-i \int_\g \d a + \frac{1}{2} \d\zeta  -i\int_\S 2B}, \qquad\qquad
\p \Sigma = \gamma,
\end{equation}
where $\S$ is a surface bounded by the loop $\g$ over which the flat $B$
field is integrated. The extension from the loop to the surface is unique,
because the difference between two such surfaces, $\S$ and $\S'$, is closed
and the integral amounts to a trivial phase,
\be
\label{B-dep}
2 \int _{\S-\S'} B= 2\pi n, \qquad\qquad n\in \Z\,.
\ee

The new fermion field \eqref{fermion} is a functional of all dynamical
fields in the theory, $a,\z,B$,  with the $B$ dependence leading to
additional phases $\int_{\S, \p \S=\g} B$.
However, these disappear from physical quantities, such
as fermion bilinears $\psi[\g]\psi^\dag[\g']$: letting the two
loops be equal $\g=\g'$, the equation \eqref{B-dep} again shows
cancellation. The same results apply to solitonic
states $\s_{N_0},\mu_{M_0}$, written in terms of
$\Tilde{a}, \Tilde{\z}$, provided that $(N_0,M_0)$ take the
values listed in the partition functions
\eqref{Z_B}. Indeed, these are the gauge
invariant excitations of the theory.

Note that in the odd sector
\eqref{odd-sector} the smallest values are $(N_0,M_0)=(\pm 1,\pm 1/2)$,
and therefore there are no physical states without $\sigma$'s.
Namely, $\mu_{1/2}$ solitons with half-integer value, $M_0=1/2$, always appear
in combination with $\s$, such that half--integer Aharonov-Bohm
(semion) phases $\th/2\pi=M_0=1/2$  are not observed.\footnote{
  Note also that monodromies are not affected by the presence of $B$.}

As usual, fractional monodromies are associated to fractional charges,
whose states are not globally defined. Let us argue that the fermion
field \eqref{fermion} is globally defined in spite of being composed of
a fractional charge.  In general, a Wilson loop for a charge $q$
state, $W[\g]=\exp(iq\int_\g C)$, is globally defined if it is
invariant under large gauge transformations of $C$:
$C \ra C + \lambda$, with $\d \lambda =0$ and
$\oint \lambda \in 2\pi \Z$.  We consider two open
surfaces $\S, \S'$ bounded by $\g$, $\p \S = \p \S'= \gamma$, and evaluate the
difference between $W[\S]$ and $W[\S']$
\begin{equation}
  \label{global-W}
W[\S] W^\dag[\S'] = e^{i q \int_{\Sigma-\Sigma'} \d C} = e^{2 \pi i N q},
  \qquad \int_{\Sigma-\Sigma'} \d C = 2 \pi N, \quad N\in\Z\,.
\end{equation}
In the last step we used the Dirac quantization condition for $C$ on the
closed surface $\Sigma-\Sigma'$. Thus, Wilson loops 
are invariant under large gauge transformations if $q \in \Z$.
This is how the Dirac condition manifests itself on these operators.

We now check the condition \eqref{global-W} for the fermion
field (\ref{fermion}) and find
\begin{equation}
  \psi[\Sigma] \psi^\dagger[\Sigma'] =
  e^{-i \int_{\Sigma-\Sigma'} \d a + 1/2 \d \zeta + 2B} =
  e^{- 2 \pi i (1/2 + 1/2 +1)} =1,
\end{equation}
using $\int B = \pi$ in the odd sector, $\int \d a = 2 \pi M_0=\pi$
and $\int \d \z = 2 \pi N_0=2\pi$. This shows that the fermion excitation
\eqref{fermion} is globally defined, as advertised.

In presence of the $B$ field, the charge operators
\eqref{boundary-charges} should also be rewritten in terms of the
gauge invariant fields $\Tilde{a}, \Tilde{\z}$, but their commutation
relations with the charged fields \eqref{boundary-charges} are
unchanged
\begin{align}
\label{tilde-charges}
Q& = \frac{1}{2 \pi} \int_{S^2}\d \Tilde{\zeta}, \qquad\qquad\quad
  Q_T=\frac{1}{2 \pi} \int_{S^2}\d \Tilde{a},
\\
\left[Q, \sigma_{N_0} \right] &=
\frac{N_0}{k} \sigma_{N_0},
\qquad\qquad \left[Q_T, \mu_{M_0} \right] =
\frac{M_0}{k} \mu_{M_0}\,.
\end{align}

Let us finally remark that, when considering the whole system with 
bulk and boundary, the fermion
(\ref{fermion}) is not gauge invariant because of the gauge
transformation $b \to b + \d \lambda$, $\zeta \to \zeta +
\lambda$. The extended bulk-boundary gauge invariant fermion is therefore
\begin{equation}
  \psi[\gamma, \Sigma] = \psi[\gamma] e^{\frac{i}{2} \int_\Sigma b},
  \qquad \p \Sigma = \gamma\,,
\end{equation}
where $\Sigma$ is a surface in the bulk attached to the boundary loop
$\gamma$. This generalization can be interpreted with the fact that
the boundary theory requires the bulk contribution to be well--defined
(gauge invariant and anomaly free). This theme has also been discussed
in section 3.2.1.

%-4.4----------------------------------------------
\subsection{Bulk theory in presence of the $B$ field}

In previous analyses we remarked that the introduction of the $B$ field
leads to the extra term
\eqref{anomaly-action} in the bulk response action, with respect to
the original expression \eqref{response-action},
\begin{equation}
\label{bulk-anomaly-AB}
S_{ind}[A,B] = i\int_{\M_4} \frac{1}{8\pi}\d A\d A  -
\frac{1}{\pi} B^2.
\end{equation}
The two terms have the similar role of canceling
the two boundary anomalies, namely the U(1) parity anomaly and the
$\Z_2^{(1)}$ anomaly. The difference is that $A$ is a background,
while the flat $B$ field is dynamical, because
the $\Z_2^{(1)}$ symmetry is gauged.

We can nevertheless write a hydrodynamic bulk action in presence of both
$A,B$ fields whose integration reproduces the response action
\eqref{bulk-anomaly-AB}, in the same way as integrating 
\eqref{bulk-action} leads to \eqref{response-action}.
The derivation uses the gauge invariance under
$U(1)_\zeta \times U(1)_a$ ($U(1)_\zeta$ is the same as $U(1)_b$ in
the bulk). By applying the same reasoning used to derive
(\ref{top-surface-action}) in reverse, starting from the
boundary action with the
electromagnetic coupling (\ref{loop-action-full}),
we are led to the following expression:
\begin{equation}
\label{bulk-action-AB}
S_{bulk} = \frac{i}{2\pi}\int_{\M_4}  b \d a + 2 B \d a
+ b B + (b + 2B) \d A - \frac{1}{4} (\d a +B) \d A.
\end{equation}
The coupling with $A$ is given in terms of the
quantities $b+ 2B$ and $\d a +B$, gauge invariant under the bulk extension
of the $\Z_2^{(1)}$ symmetry \eqref{gauged-Z12-symm}
\begin{equation}
\label{gauged-Z21-bulk}
    a \ra a - \beta, \qquad b \ra b-2\d\beta,  \qquad B \ra B + \d \beta.
\end{equation}
One can check that by integrating out the dynamical fields $b$ and $a$
in (\ref{bulk-action-AB}) one indeed recovers the
response action (\ref{bulk-anomaly-AB}).

Notice that we can change to tilded variables as we did on the boundary
(cf. \eqref{tilde-act}), with $\Tilde{b} = b + 2B$, to rewrite
the bulk action \eqref{bulk-action-AB} into the form
\begin{equation}
\label{bulk-tilde-act}
S_{bulk} = \frac{i}{2\pi}\int_{\M_4}  \Tilde{b} \d \Tilde{a} +
 \Tilde{b} \d A  - \frac{1}{4} \d \Tilde{a} \d A - 2 B^2\,.
\end{equation}
Again, this is the original action (\ref{bulk-action}) written in
terms of the new variables, up to the anomaly term $B^2$.

The action for the complete bulk and boundary system,
(\ref{bulk-action-AB}) and (\ref{loop-action-full}), is gauge
invariant and anomaly free: indeed, it is gauge invariant (using
$\d B = 0$) under:
\begin{itemize}
\item
The $U(1)$ gauge symmetry of $b$ and $\zeta$, given by
  $b \ra b + \d \lambda$ and the analogous
  $\zeta \ra \zeta + \lambda$.
\item
The $U(1)$ gauge symmetry of $a$, which is $a \ra a + \d \alpha$.
\item
The $\Z_2^{(1)}$ gauge symmetry given by
  (\ref{gauged-Z21-bulk}) in the bulk and (\ref{gauged-Z12-symm}) on the
      boundary.
\end{itemize}
In conclusion, the expression (\ref{bulk-action-AB}) include the 
(minimal) modifications
of the original `bosonic' bulk action \eqref{bulk-action} which are
required to comply with fermionic boundary excitations.
We remark that more general 3+1 dimensional topological actions
have been discussed in the literature which can also describe bulk
fermionic degrees of freedom  \cite{kapustinthorngren2017}.

%-4.5--------------------------------------------------
\subsection{Time reversal symmetry and the B field}

We now discuss how the introduction of the $B$ field fits with the
time-reversal symmetry ${\cal T}$ of topological insulators.\footnote{
  In this section we turn back to Minkowski space to treat time
  reversal.}
Since the bulk and boundary action take the original form without
$B$ once written in terms of tilded variables 
(cf. \eqref{tilde-act}, \eqref{bulk-tilde-act}), 
these can be set to transform as the original fields too. Therefore, we assume
\begin{equation}
\label{TR-tilde}
{\cal T}: \ (\Tilde{a}_0, \Tilde{a}_i) \ra (\Tilde{a}_0, -\Tilde{a}_i),
\quad (\Tilde{\z}_0, \Tilde{\z}_i) \ra (-\Tilde{\z}_0, \Tilde{\z}_i),
\quad (\Tilde{b}_{0i},\Tilde{b}_{ij}) \ra (-\Tilde{b}_{0i}, \Tilde{b}_{ij}).
\end{equation}
The compatibility between transformations of original and tilded variables
is as follows. The map $\d a + B=  \d \tilde{a}$ is consistent
with time reversal provided that $B$ transform as a 'vector'
two--form, 
\begin{equation}
  \label{T-B}
{\cal T}:\; (B_{0i},B_{ij}) \ra (B_{0i}, -B_{ij}).
\end{equation}
The map between $b$  and $\Tilde{b}$ is instead consistent assuming that the pseudo--two--form $b$ mixes with $B$ under
${\cal T}: b \ra b +4B$.

As usual, the bulk plus boundary action is time reversal invariant,
but the individual actions \eqref{tilde-act} and \eqref{bulk-tilde-act}
are invariant up to the sign of the anomaly terms. The response
actions, respectively \eqref{response-full} and \eqref{bulk-anomaly-AB},
also change sign.

Now we turn to the action of time reversal on physical
states. According to the earlier discussion on bosonization,
the fermionic theory (\ref{Z_f-S})
has a zero--form $\Z_2$ symmetry generated by
$\int B$, which is identified with the fermion number symmetry,
\begin{equation}
\label{fermion-number}
    (-1)^F = e^{i\int_{S^2}B}.
\end{equation}
As said, this is consistent with the two sectors in the theory
having, respectively, bosonic states (\ref{even-sector}), with $\int B = 0$, 
and fermionic states (\ref{odd-sector}),  with $\int B = \pi$.
For fermionic topological insulators, time reversal transformations obey
${\cal T}^2 = (-1)^F$; therefore, in our description
(\ref{fermion-number}), we should have
\begin{equation}
\label{T-square}
    {\cal T}^2 = e^{i\int_{S^2}B}.
\end{equation}

The action of time reversal on states consistent with \eqref{T-square}
is achieved as follows. We consider products of $\sigma$ and $\mu$ fields
(\ref{sigma-mu-op}), in their closed loop version,
with allowed values of $(N_0,M_0)$ in the two sectors of the fermionic
theory \eqref{Z_B},
\begin{equation}
  \label{gen-stat}
\Phi = e^{-i N_0 \int \Tilde{a}-i M_0 \int \Tilde{\z}}.
\end{equation}
A naive action of time reversal is obtained by using
the field transformations \eqref{TR-tilde} in \eqref{gen-stat} and
antiunitarity, leading to 
\begin{equation}
  {\cal T}\Phi  {\cal T}^{-1}\sim   e^{-i N_0 \int \Tilde{a}+ iM_0 \int \Tilde{\z}}.
\end{equation}
This cannot be correct, because it obeys ${\cal T}^2=1$.

Therefore, we are led to add a non-trivial $B$-dependent phase 
factor to the transformation, as follows:
\be
\label{T-states}
  {\cal T} \Phi  {\cal T}^{-1} \equiv e^{i\int_{S^2}\frac{B}{2}} \,
e^{-i N_0 \int \Tilde{a} +i M_0 \int \Tilde{\z}}.
\ee
We remark that the phase is trivial, $\int B = 0$, for all bosonic
states in the even sector of the theory, thus ${\cal T}^2=1$.
In the odd sector,  $\int B = \pi$, the phase is not vanishing
and furthermore it obeys:
${\cal T} \exp(i\int_{S^2}B/2) {\cal T}^{-1}= \exp(i\int_{S^2}B/2)$
(cf. \eqref{T-B}). We find,
\begin{equation}
\label{T2-odd}
{\cal T}^2\Phi  {\cal T}^{-2}={\cal T}\left(e^{i\int_{S^2}\frac{B}{2}}
  e^{-i N_0 \int \Tilde{a} +i M_0 \int \Tilde{\z}}\right) {\cal T}^{-1}=
  e^{i\int_{S^2}B} \Phi= (-1)^F \Phi .
\end{equation}
In conclusion, the definition \eqref{T-states} of
time-reversal transformations correctly realizes 
${\cal T}^2= (-1)^F$.

\bigskip

{\bf Bosonized fermions and Dirac spinors.}
The standard presentation of 2+1d Dirac fields in terms
of bi-spinors and Pauli matrices is a reducible representation
of the Poincar\'e group. Indeed, spin is a pseudoscalar and irreducible
representations are one dimensional
\cite{binegar1982poincare,jackiwnair1991poincare}. In the massless case,
the spinors decompose in two irreducible one-dimensional representations
with quantum numbers of momentum and the 2+1d analog of helicity, which
are solutions of the Dirac equation.
Time reversal transformations are realized on spinors by
${\cal T}=\eta \s_2 K$, where $\eta$ is a phase, $\s_2$ a Pauli matrix
and $K$ complex conjugation. Naturally, the two 
one-dimensional Poincar\'e representations
map one into another under ${\cal T}$.

In the bosonization language of this paper, these two one-dimensional
irreducible representations should be identified with the fermion
\eqref{fermion} and its time-reversal partner given by
\eqref{T-states}.

%-4.6--------------------------------------------------------
\subsection{Fermionic theory from the $k=2$ loop model and
  the toric code }

In this section we describe the alternative derivation of the
fermionic theory from the bosonic loop model with $k=2$.
Its topological aspects, given by the $k=2$ BF theory, are shared with
the toric code, the bosonic topological phase of matter
which was mentioned in section 4.2.1.

Our first observation is that the fermionic partition functions
$Z_f^{NS}, Z_f^{\wt {NS}}$ \eqref{Z_f-k2} obtained in section 4.3.2
resemble the loop-model bosonic $Z$ for $k=2$ \eqref{loop-model-Z},
hereafter denoted $Z_b[k=2]$. Actually, the solitonic spectrum is the 
same, although a parity rule in the summations eliminates half of the
states (and there is the $(-1)^F$ sign).

Let us compare the fermionic spectrum \eqref{Z_f-k2}
with those occurring in the $k=2$ and $k=1$ bosonic theories.
\begin{itemize}
\item
On one hand, $Z_f[\eta_\pm]$ contains a subset of the $k=2$
  spectrum in $Z_b[k=2]$, owing to the parity rule, $n_0=m_0$ mod $2$.
\item
On the other hand, $Z_f[\eta_\pm]$ involves states
  that are not present in the $k=1$
  bosonic theory $Z_b[k=1]$ \eqref{loop-model-Z}, \textit{i.e.}
  those with half-integer $M_0$, and a parity rule is also present.
\end{itemize}
These facts suggest that the fermionic theory is related to
both the $k=1$ and $k=2$ bosonic loop models by slightly different 
bosonic-fermionic maps.
The $k=2$ map is simpler for the fact that the needed fermionic states are
already present in the bosonic theory, but other aspects are troublesome:
\begin{itemize}
\item Topological insulators within the
  classification of SPT phases are associated to systems with
  short-range entanglement and no topological order, as the
  integer Hall effect \cite{wen2017colloquium,ryu2016classification}.
  These properties fit with the $k=1$ bulk BF
  theory \eqref{bulk-action} and the corresponding response action
  \eqref{response-action}, which can be easily derived from massive
  free fermions \cite{cappellianomaliescondmat}.
\item
  On the contrary, the toric code is a $\Z_2$ lattice gauge theory
  which possesses semion excitations $e,m$ with fractional charge and
  fractional monodromies, and corresponding nontrivial topological order.
  Note that the $k=2$ BF theory also describes the low-energy
  limit of the Abelian Higgs model of superconductivity
  \cite{hansson2004superconductors}.
\end{itemize}
Nonetheless, with hindsight gained from the previous analysis, we shall 
find that both unwanted features, that of fractional statistics, and
topological order, disappear in going from the $k=2$ bosonic model to the
fermionic theory. This is the consequence of gauging the $\Z_2^{(1)}$ symmetry.

Let us now outline the fermionization of the $k=2$ bosonic theory,
paralleling the analysis in section 4.3.
The full action in 2+1 dimensions is
\begin{equation}
\label{k2-boundary-action}
\begin{split}
  S =&  \frac{i}{2\pi}\int_{\M_3} 2 \zeta \d a +
  2\left(a +\zeta \right) B + 2 (\d \zeta
  +B) A - \frac{1}{4}(\d a + B)A +
  \\
  & + \frac{g_0}{8\pi}\int
  \left(f +B \right)_{\mu\nu} \frac{1}{\p}  \left(f +B\right)^{\mu\nu}
  \qquad\qquad (f=da),
\end{split}
\end{equation}
including the $A$ background and the flat $\Z_2$ two--form field $B$.
Note that the factor of $2$ in the electromagnetic coupling $d\z A$ is
not standard: it redefines the measured charge of `electric' semions.
 Note also that the $k=2$ action \eqref{k2-boundary-action} is formally equal to
its $k=1$ counterpart \eqref{loop-action-full} by replacing $\z\to \z/2$.

As described in section 4.2.1, the theory \eqref{k2-boundary-action}
possesses a $\Z^{(1)}_2\times \Z^{(1)}_2$ global symmetry, with currents
$*a$ and $*\z$. The diagonal
subgroup is gauged by coupling its current $*(a+\z)$ to the dynamical
$B$ field, as in \eqref{diag-symm}.
The gauge invariance is \eqref{Z2-gauge}, and the action develops
the anomaly term \eqref{toric-code-anomaly}. After removing this
anomaly from the bulk, the fermionic partition function
\eqref{Z_f-k2} is obtained by gauging the bosonic partition
function according to \eqref{gauged-Z}.

Let us find the constraint on charge quantizations due to the summation
over the $B$ field, the analog of conditions \eqref{N0-B} and \eqref{M0-B}
in the $k=1$ case. The 
equations of motion for \eqref{k2-boundary-action} (with $A=0$),
$B = -\d \z$ and $B = -\d a$, can be integrated, leading to the conditions
\be
\label{k2-sectors}
\begin{split}
& - \frac{1}{2\pi} \int_{S^2} B =\frac{1}{2\pi}\int_{S^2} d\z=Q=
  \frac{ n_0}{2}=0,\frac{1}{2}, \quad {\rm mod}\ 1,
  \\
& - \frac{1}{2\pi} \int_{S^2} B =\frac{1}{2\pi}\int_{S^2} da =Q_T=
\frac{ m_0}{2}=0,\frac{1}{2},\quad {\rm mod}\ 1,
\end{split}
\ee
using the definitions \eqref{boundary-charges}.
As in the $k=1$ case, the solutions identify two sectors:
in the even one, for $\int B = 0$ mod $2\pi$,  $n_0$ and $m_0$ are even,
while in the odd sector, with $\int B = \pi$ mod $2\pi$, 
both $n_0$ and $m_0$ are odd. 
Thus we reobtain the parity rule present in
the fermionic partition function \eqref{Z_f-k2}.
This completes the map between the bosonic $Z_b[k=2]$ and fermionic 
$Z_f[\eta_\pm]$ partition functions.

Now we turn to the representation of the fermion field. It is given by the product of electric and magnetic
semions $e,m$ with charges $n_0=m_0=1$,
\be
\label{alt-fermion}
\psi[\g]= :e\, m:=e^{-i\int_\g \Tilde{a}+ \Tilde{\z}}
  =e^{-i\int_\g a+ \z -2i\int_\S  B},
  \ee
  where the gauge invariant tilded variables are now
  $\d \Tilde{a}=\d a +B$, $\d \Tilde{\z}=\d \z +B$. 
Clearly, the odd and even sectors contain corresponding numbers of
electrons. The odd sector acquires the sign $(-1)^F=-1$ in the
partition function $Z_f^{\wt{NS}}$ in \eqref{Z_f-k2}.

Let us see how the statistics of fermions is related to the monodromy
of semions $e,m$ and the effects of gauging the $\Z_2^{(1)}$ symmetry.
The monodromy phase between excitations $(n_0,0)$, $(0,m_0)$
is $\Th/2\pi=n_0 m_0/2$: between
fermions $\psi=$ :$em$:, one correctly obtains $\Th/\pi=1/2+1/2=1$.
Note that basic semion excitations, with $(n_0,m_0)=(1,0), (0,1)$,
would lead to fractional values, but they are absent
from the spectrum \eqref{Z_f-k2}, being forbidden by
$\Z_2^{(1)}$ gauge invariance.
Therefore, anyons of the original $k=2$ bosonic theory
disappear in the fermionic theory by gauging.

Note also that the monodromy of all $(n_0,m_0)$ excitations in \eqref{Z_f-k2}
with the fermion is trivial, because
$\Th/2\pi=(n_0+m_0)/2=1$, owing to the parity rule $n_0=m_0$ mod $2$.
In particular, spatial fermion loops  $\psi(\g)$ considered as topological
defects are actually invisible to all excitations.
The `fermion condensation' picture associated to the gauging
of the $\Z_2^{(1)}$ symmetry  is indeed realized as advertised.

The physical electric charge of the electron is given by twice the
charge of semions $Q$, owing to the mentioned normalization of the
coupling $2 A d\z= A_\mu J_{EM}^\mu$ in \eqref{k2-boundary-action},
leading to
\begin{equation}
  \label{k2-charge}
  Q_{EM} = \int_{S^2} * J_{EM} = \frac{2}{2 \pi} \int_{S^2} \d \Tilde{\z}
  = 2 Q= 2 \frac{n_0}{2} \in \Z\,,
\end{equation}
in both sectors  (recall also
the field commutator $[Q,\psi(x)]=\psi(x)/2$
\eqref{boundary-charges-comm}).
The magnetic charge is again $1/2$.

We remark that the normalization \eqref{k2-charge} of the physical charge
also implies
standard fermionic parity anomaly \eqref{response-action}, with $k=1$, upon integration
of the action \eqref{k2-boundary-action} over all dynamical fields. 
Actually, the same occurs in the bulk BF theory \eqref{bulk-action}, by
considering the more general electromagnetic coupling parameterized by $n$,
\begin{equation}
\label{general-bulk-action}
S_{bulk} = i \int_{\M_4} \frac{k}{2 \pi} b \d a +
\frac{n}{2\pi} b\d A - \frac{1}{8\pi} \d a \d A.
\end{equation}
Here $k,n \in \Z$ for gauge invariance, while we have already fixed
$\theta = \pi$. When $k=n$, integrating out $a$ and $b$
yields the correct topological insulator electromagnetic response
\be
S_{ind}[A] = \frac{i}{8\pi} \int_{\M_4} \d A \d A.
\ee
From the point of
view of getting this result, the BF coupling $k=2$ also works
provided that we take $n=k=2$.

We now discuss the fate of the topological order of the bulk theory 
\eqref{general-bulk-action} for $k=2$ (setting $A=0$).
Let us consider the manifold $\M_4=S^2\times S^1\times \R$,
for convenience. In Hamiltonian formulation, 
the topological order is equal to the dimension of the representation
of the loop and surface operators algebra
\be
\label{UW-algebra}
W=e^{i\int_{S^1} a}, \qquad U=e^{i\int_{S^2} b}, \qquad\qquad
WU=e^{i\pi} UW,
\ee
(given by $2$).
These express the $\Z^{(1)}_2\times \Z^{(1)}_2$ global symmetry, in which
$U$ is the symmetry generator and $W$ is the charged state for one
subgroup and viceversa for the other one.
The gauging by the $B$ field leads to the action (cf.
\eqref{k2-boundary-action} as well as \eqref{bulk-action-AB} with $\z \to 2\z$)
\be
S_{bulk} = i \frac{2}{2 \pi}\int_{\M_4}  b \d a + (\d a+ b)B\,,
\ee
where actually the term $\d a B$ vanishes (for vanishing $a$ values at
$t=\pm\infty$). We are therefore effectively gauging the $\Z_2^{(1)}$ symmetry generated by $U$.

In this equation, the integration over $B$ is a discrete Gauss law:
 physical states should be annihilated by the symmetry generator $U$  in
\eqref{UW-algebra}.
From the representation of the loop algebra on the two-dimensional
ground state manifold $|\W_\pm\rangle$,
\be
W|\W_\pm\rangle=\pm |\W_\pm\rangle, \qquad U |\W_\pm\rangle= |\W_\mp\rangle,
\ee
we can easily conclude that there is a single gauge invariant
state, $| phys \rangle= |\W_+ \rangle + |\W_- \rangle $, which
is left invariant by $U$. Therefore, the gauging by $B$ 
leads to a single ground state, \textit{i.e.} it
eliminates the topological order. This is expected, since the topological order can be seen as the spontaneous symmetry breaking of the $\Z_2^{(1)}$ symmetry that we are gauging \cite{gaiotto2015generalized,mcgreevy2023generalized}.

This completes the map from the $k=2$ bosonic loop model
to the fermionic theory. We have found that there are two
bosonic theories which lead to the same fermionic theory after gauging
the characteristic one-form symmetry.

A final remark concerns the fate of the duality symmetry obeyed by the
loop model.  The fermionic partition function $Z_f[\eta_\pm]$
\eqref{Z_f-k2} is not self-dual under the $k=1$ transformation of the
original theory \eqref{loop-model-duality}, owing to the additional
states with half-integer $M_0$ values.  $Z_f[\eta_\pm]$ is instead
invariant for the $k=2$ transformation:
\begin{equation}
    g_0\ \lra\ \frac{4}{g_0}, \qquad n_0 \ \lra\  m_0,
\end{equation}
because bosonic \eqref{loop-model-Z} and fermionic \eqref{Z_f-k2}
spectra only differ for the parity rule, which is invariant under
$n_0 \lra m_0$.

More importantly, the fermionic theory is self-dual under a variant of
the particle--vortex map which holds between fermionic theories
\cite{metlitski2016,tongwebofduality}.  It is defined on the action
\eqref{k2-boundary-action}, conveniently rewritten in terms of tilded
variables, by replacing the $A$ background with the dynamical field
$ c$ and adding a BF coupling $c \d A /4\pi$, having $1/2$
normalization.\footnote{ Note that this transformation is not entirely
  correct, because the $1/2$ coupling is not actually gauge
  invariant. A more refined definition is found in section 4 of
  \cite{wittenwebofduality}.  Much like as in the FQHE case
  \eqref{FQHE}, one should add a new dynamical gauge field such that
  everything is gauge invariant. This extra field cannot be integrated
  out because of inconsistent flux quantization. However, integrating
  it out naively yields the duality stated in
  \cite{metlitski2016,tongwebofduality}, which is sufficient for our
  purposes.}
Integration over $c$ leads to an equivalent action with
$\Tilde{a} \lra \Tilde{\z} $ and $g_0\ \lra\ 4/g_0$. As a matter of
fact, the bosonic $k=2$ and fermionic dualities are effectively the
same transformation. The self-dual value is $g_F\equiv g_0/2=1$: this
mismatch by a factor of two is also found in the self-dualities of
mixed-dimension QED$_{4,3}$ with fermionic and bosonic matter
\cite{sonmixedqed,sonmixedqedscalar}, which correspond to the loop
model for large number of flavors.

%-5-----------------------------------------
\section{Conclusions} 
\label{conclusions}

In this paper, we developed the analysis of the effective field theory
of 3+1 dimensional topological insulators, based on the bulk BF action
\cite{BFTI} and the boundary loop model providing the dynamics
for surface excitations \cite{loopmodel}.
Among the interesting features of this model motivating our study, we
remark the presence of a critical line with conformal invariance, and of
solitonic states with electric and magnetic charges $N_0$ and $M_0$,
whose spectrum of scaling dimensions is known exactly.

We showed that the solitonic local operators $\s(x), \mu(x)$
can be represented by open Wilson lines of the two gauge fields $a,\z$,
but also in terms of monopoles of the dual field,
employing results of Refs. \cite{marino,frohlich87,frohlich95}.
These identifications are exact in the topological limit of the
loop model, where the interaction strength $g_0$ vanishes.
For $g_0\neq 0$, we computed the soliton correlation functions and verified their conformal invariance, with
scaling dimensions in agreement with the results of the partition function.
Actually, the derivation is simpler in terms of the dual monopoles,
while that using Wilson lines requires some assumptions on the
dynamics of fluctuating loops at criticality.
We then defined the fermion field $\psi$ as the product  $\s \mu$ with
charges $(N_0,M_0)=(1,1/2)$.

The bosonization of 2+1 dimensional fermions not only requires
the identification of the fermion field, but also of other properties, such as the fermion number $\Z_2$ symmetry $(-1)^F$,
the spin sectors and the time reversal symmetry obeying
${\cal T}^2=(-1)^F$. These features have been implemented in the bosonic loop model by using
the construction of
Refs. \cite{gaiottokapustinspinTQFT1,gaiottokapustinspinTQFT2}.
This was originally formulated for bulk topological theories
on triangulated manifolds, but can be extended to the continuum
and for boundary theories, which include a topological part
and an interaction term (for $g_0\neq 0$).

The map leading from the bosonic to the fermion theory
starts from the identification of the $\Z_2^{(1)}$
one-form symmetry generated by fermion loops, and its universal
anomaly. Once this anomaly is canceled by the bulk, the symmetry
is gauged by a coupling to a dynamical $\Z_2$ flat two-form field $B$.
The following results follows:
\begin{itemize}
\item The $B$ field implements the half-integer magnetic charge of the fermion, and
  divides the bosonic spectrum into sectors. The $B$ field also generates
  the fermion number symmetry and enters in the definition of time reversal
  transformations.
\item The gauging of $\Z_2^{(1)}$ symmetry determines the spin sectors
  of the fermionic theory. Furthermore, physical states are selected
  by the requirement of $\Z_2^{(1)}$ gauge invariance.
\end{itemize}

The new ingredients of $\Z_2^{(1)}$ gauge symmetry and $B$ field
are very important for the correct definition of a fermionic theory and thus fermionic topological phases. The standard effective theory based on
the BF theory is not sufficient and only applies to bosonic topological phases.
Actually, we showed that two BF theories, with $k=1$ and $k=2$ couplings,
are mapped into the same fermionic theory. 

The fact that the BF theory by itself is not sufficient for describing a fermionic topological phase is actually expected on general grounds: it is believed that while bosonic phases have a TQFT effective description, the low energy description of a fermionic phase is a spin TQFT \cite{wen2017colloquium,gaiottokapustinspinTQFT2,kapustinthorngren2017}. In this work, we made explicit this difference by using bosonization methods to show how the theory partition function (and thus the physical states) changes under the fermionization map.

A related aspect is the interplay between fermionization of
bulk and boundary theories, needed for accommodating respective
fermionic excitations, higher-form $\Z_2$ symmetries
and anomalies.  In this work, we studied the map for
the boundary theory and did not delve into the proper definition of
bulk fermions.

The 2+1 bosonization discussed in this work can be further
investigated by analyzing other aspects of the fermionic theory and
other geometries.  The bosonization of 3+1 dimensional relativistic
fermions is also very interesting for its potential application to
nonperturbative physics of fundamental interactions, and can
be similarly formulated by considering 4+1d
topological theories and topological insulators \cite{hydroAC}.

{\bf Acknowledgements} We would like to thank A. G. Abanov,
P. B. Wiegmann and G. R. Zemba for interesting exchanges on the topics
of this work.  The work of A.C. has been partially supported by the
grants PRIN 2017 and PRIN 2022 provided by the Italian Ministery of
University and Research.

\appendix 
%-A-----------------------------------------------------
\section{Computation of soliton correlation function}
\label{app-A}
In this appendix we explicitly compute the correlator
(\ref{sigma-corr}). This amounts to the evaluation of the action
(\ref{loop-model-zeta}) on the monopole solution $ f(x_1) - f(x_2)$,
with $f$ given in (\ref{monopole}).  We thus have:
\begin{equation}
    \braket{\s_{N_0}(0) \s_{N_0}^\dag(R)} = e^{-S(0,R)},
\end{equation}
with 
\begin{equation}
\begin{split}
  &S(0,R) \equiv S[N_0(f(0)-f(R))] = \frac{k^2}{16 \pi^3 g_0}
  \int \d x^3 \d y^3 F_{\mu \nu}(x) \frac{1}{(x-y)^2} F_{\mu \nu} (y), \\
  &\ F_{\mu \nu}(x)  =
  \frac{N_0}{2 k} \epsilon_{\mu \nu \rho} \left( \frac{x_\rho}{|x|^3}-
    \frac{(x-R)_\rho}{|x-R|^3}\right).
\end{split}
\end{equation}
We neglect the `self--interaction terms' ($f(0)f(0)$ and $f(R)f(R)$)
that arise in evaluating $S(0,R)$: they are not
relevant for determining how the correlator 
depends on $R$ and can be reabsorbed by a
normalization of $\s(x)$. The object to evaluate is then
\begin{equation}
  S(0,R) = - \frac{N_0^2}{16 \pi^3 g_0}
  \int \d x^3 \d y^3 \frac{x^\mu (y-R)_\mu}{|x|^3 |x-y|^2 |y-R|^3}\,.
\end{equation}
This dimensionless integral is both infrared and ultraviolet divergent:
the infrared cutoff is the system size $L$ and the ultraviolet
length is roughly the monopole distance $R$. Thus we expect the
result $S=-c \log(L/R)$ and need to determine the coefficient.
Using 
\[ \p_\mu \frac{1}{|x|} = - \frac{x_\mu}{|x|^3},\] and integrating by
parts, we obtain
\begin{equation}
  S(0,R) = \frac{N_0^2}{8 \pi^3 g_0} \int \d x^3 \d y^3
  \frac{1}{|x| |x-y|^4 |y-R|}.
\end{equation}
Denominators can be exponentiated by using the Gamma function identity
\begin{equation}
\frac{1}{x^z} =  \frac{1}{\Gamma(z)} \int_0^{\infty} \d t t^{z-1} e^{-xt},
\end{equation}
leading to
\begin{equation}
  S(0,R) = \frac{N_0^2}{8 \pi^4 g_0} \int \d^3 x \d^3 y \int_0^{\infty}
  \d t \d u \d s\, t^{-1/2} u^{-1/2} s\, e^{-x^2t-(y-R)^2 u -(x-y)^2 s}.
\end{equation}

The integrals over $x$ and $y$ are now Gaussian and yields:
\begin{equation}
  \begin{split}
  S(0,R) &= \frac{N_0^2}{8 \pi g_0} \int_0^{\infty} \d t \d u \d s
  \frac{t^{-1/2} u^{-1/2} s}{(ut +ts +su)^{3/2}}
  \exp \{ -R^2 \frac{uts}{ut +ts + su} \} = \\
  & = \frac{N_0^2}{8 \pi g_0} \int_0^{\infty} \d t \d u
  \frac{t^{-1/2} u^{-1/2}}{(ut +t +u)^{3/2}} \int_0^{\infty} \frac{\d s}{s}
  \exp \{ - s R^2 \frac{ut}{ut +t + u} \},
 \end{split}
\end{equation}
where in the second expression we changed variables $u=s u'$,
$t=st'$ and renamed again $u',t'$ as $u,t$.
The integral over $s$ is divergent for $s \ra 0$: this 
is the infrared divergence which we regularize by requiring $s> 1/L^2$,
given that $s$ has mass dimension 2. We find
\begin{equation}
  \label{s-int}
\begin{split}
  \int_{1 /L^2}^{\infty} \frac{\d s}{s} e^{-s R^2 A}=
  E_1(R^2A/L^2) &= -\gamma - \log(R^2A/L^2) + O(R^2/L^2), \\
  A &\equiv \frac{ut}{ut +t + u}\,,
\end{split}
\end{equation}
where $E_1(z)$ is the exponential integral function with the following
expression  off the negative real axis:
\begin{equation}
    E_1(z) = -\gamma - \log(z) - \sum_{k=1}^\infty \frac{(-z)^k}{k (k!)}.
\end{equation}
Inserting \eqref{s-int} in the full expression of $S(0,R)$ yields
\begin{equation}
\begin{split}
  S(0,R) = - \frac{N_0^2}{8 \pi g_0} \int_0^{\infty} \d t \d u
  \frac{t^{-1/2} u^{-1/2}}{(ut +t +u)^{3/2}}
  \{ \gamma + \log(R^2/L^2) + \\
  +  \log\left(\frac{ut}{ut +t +u}\right) + O(R^2/L^2) \}.
\end{split}
\end{equation}
Only the term proportional to $\log(R^2)$ is
actually meaningful: higher order contributions $O(R^2/L^2)$
can be neglected in the limit $L\ra \infty$, while the other
two terms are independent from $R$ and can be reabsorbed
in field redefinitions. We are thus left to
compute
\begin{equation}
  S(0,R) = - \frac{N_0^2}{8 \pi g_0} \log(R^2/L^2)
  \int_0^{\infty} \d t \d u \frac{t^{-1/2} u^{-1/2}}{(ut +t +u)^{3/2}}. 
\end{equation}
The remaining integral
\begin{equation}
  \int_0^{\infty} \frac{\d u}{u^{1/2}} \int_0^{\infty} \d t
  \frac{1}{(t(ut +t +u)^{3})^{1/2}} = 2\int_0^{\infty}
  \frac{\d u}{u^{3/2}(1+u)^{1/2}} =-4,
\end{equation}
is divergent for $u\to 0$ as $\int du/u^{3/2}$ (UV divergence)
and can be computed by analytic continuation
(or by inserting a cut-off and keeping the finite part).

The final result is
\begin{equation}
    S(0,R) = \frac{N_0^2}{2 \pi g_0} \log(R^2/L^2)\,,
\end{equation}
and therefore 
\begin{equation}
\label{final-correlator}
\braket{\s_{N_0}(0) \s_{N_0}^\dag(R)} = e^{-S(0,R)} =
\left( \frac{L}{R} \right)^{2\D_\s}, \qquad
\D_\s= \frac{1}{2\pi} \frac{N_0^2}{g_0}.
\end{equation}
This completes the derivation of the $\sigma$ correlation 0function. 

\bibliographystyle{JHEP}
\bibliography{dirac1.bib}

\end{document}